\documentclass[namedreferences,optionalrh]{spr-sola}
\usepackage{graphicx}        
\usepackage{amssymb}        
\usepackage{amsmath}
\usepackage{color}           

\usepackage[colorlinks=true,citecolor=blue]{hyperref}

\usepackage{comment}

\usepackage[margin=1.0in, bottom=1.75in, top=1.0in]{geometry}

\newcommand{\bm}[1]{{\mbox{\boldmath $#1$}}}

\chardef\us=`\_

\begin{document}

\begin{frontmatter}

    \title{Recent Advances in Numerical Modeling of Solar Inertial Modes}
    
    \author[addressref={mps},corref,email={bekki@mps.mpg.de}]{\inits{Y.}\fnm{Yuto}~\lnm{Bekki}\orcid{0000-0002-5990-013X}}

    \author[addressref={mps,goettingen}]{\inits{L.}\fnm{Laurent}~\lnm{Gizon}\orcid{0000-0001-7696-8665}}

    \runningauthor{Bekki \& Gizon}
    \runningtitle{Numerical Modeling of Solar Inertial Modes}
    \address[id=mps]{Max-Planck-Institut f{\"u}r Sonnensystemforschung, Justus-von-Liebig-Weg 3, 37077 G{\"o}ttingen, Germany}
    \address[id=goettingen]{Institut f{\"u}r Astrophysik und Geophysik, Georg-August-Universit{\"a}t G{\"o}ttingen, Friedrich-Hund-Platz 1, 37077 G{\"o}ttingen, Germany}

\begin{abstract}

Inertial modes are low-frequency global oscillations in rotating fluids, restored by the Coriolis force. 
Long-term observations of horizontal flows on the solar surface  have led to the detection of various inertial modes on the Sun, opening a new window into the structure and dynamics of the solar interior.
Here, we focus on recent advances in the theoretical modeling of solar inertial modes within a purely hydrodynamic framework, leaving magnetohydrodynamic effects beyond the scope of this review.
Because the low-frequency oscillations of a differentially rotating spherical shell depend sensitively on the rotation profile and the properties of the medium, they are primarily studied using numerical models.
We first summarize how linear eigenmode analyses have enabled the identification and classification of different families of inertial modes observed on the Sun, and how these models reveal the strong diagnostic potential of inertial modes for probing poorly constrained parameters in the deep convection zone, such as turbulent viscosity and superadiabaticity.
The amplitudes of linearly stable modes can be estimated from their response to stochastic forcing, whereas those of linearly unstable modes must be determined using nonlinear numerical modeling. 
We further discuss the possible dynamical roles of large-amplitude inertial modes in transporting heat and angular momentum in the Sun. 
By clarifying what has been established and what remains uncertain, this review provides a framework for interpreting current observations and guiding future theoretical developments in this rapidly evolving field.

\end{abstract}


\end{frontmatter}


\section{Introduction} \label{sec:intro}

Inertial modes are low-frequency global oscillations in rotating fluids whose restoring force is the Coriolis force \citep[][]{greenspan1968}. Although their existence in the Sun had long been predicted theoretically \citep[e.g.,][]{provost1981,saio1982,unno1989}, their detection requires long-term observations spanning many years. This observational challenge was overcome thanks to long-duration time series provided primarily by the Helioseismic and Magnetic Imager (HMI) on board the Solar Dynamics Observatory (SDO, since 2010) and by the Global Oscillation Network Group (GONG, since 2001).

The first unambiguous detection of solar inertial modes was reported by \citet{loeptien2018}, based on six years of SDO/HMI observations that revealed equatorial Rossby modes.
In astrophysics, Rossby modes (or r modes) are traditionally defined as a special class of inertial modes characterized by quasi-toroidal motions, for which radial motions are substantially weaker than the horizontal motions \citep[][]{papaliozou1978,provost1981,rieutord2001}.
\citet{loeptien2018} found that these modes appear as excess power in the sectoral ($\ell=m$) component of the near-surface radial vorticity spectrum and follow the dispersion relation of classical Rossby modes in a uniformly rotating sphere, $\omega = -2m\Omega_{\rm eq} / [\ell(\ell+1)]$, where $\omega$ is the frequency measured in the rotating frame, $m$ is the azimuthal order, $\ell$ is the spherical harmonic degree, and $\Omega_{\rm eq}$ is the equatorial rotation rate of the Sun. 
We adopt the convention that negative frequencies correspond to retrograde propagation for positive values of $m$.
No additional power ridges consistent with the dispersion relations of non-sectoral Rossby modes have been identified in the observations. 
The detected solar Rossby modes are characterized by north–south symmetric radial vorticity patterns that peak at the equator, and thus are conventionally called \textit{equatorial} Rossby modes. 
They were observed over the azimuthal order range $3 \leq m \leq 15$, with typical oscillation periods of several months and near-surface velocity amplitudes of approximately $1$–$2$~m~s$^{-1}$.

Building on the work by \citet{loeptien2018}, \citet{gizon2021} have demonstrated that equatorial Rossby modes represent only a subset of a much richer spectrum of solar inertial modes. 
Unlike the Rossby modes, whose quasi-toroidal nature allows for analytical solutions under the assumption of uniform rotation, general inertial modes in a spherical shell cannot be described by simple analytical solutions. 
In particular, their spatial patterns are not represented by spherical harmonics. 
To overcome this difficulty, \citet{gizon2021} analyzed the latitude-dependent power spectra at fixed azimuthal order $m$, and detected dozens of additional inertial modes based on their power excess seen over a wide range of latitudes at a single frequency.
The detected modes were further compared with the linear model of solar inertial modes \citep[][]{bekki2022a}, which enabled systematic identification and classification of the observed inertial modes.
The newly detected and identified modes include high-latitude modes, characterized by spiral flow patterns around the poles, and critical-layer modes, whose eigenfunctions are concentrated near the viscous critical layers where the mode phase speed matches the local differential rotation \citep{gizon2020}. 
Although the critical-layer modes tend to have similar mode amplitudes to those of the equatorial Rossby ($\sim 1$~m~s$^{-1}$), the high-latitude modes are found to have significantly larger amplitudes.
In fact, the $m=1$ high-latitude mode has the mode amplitude of $\sim 10$~m~s$^{-1}$ near the surface, which is the largest among all the observed inertial modes.
\citet{liang2025} note that this particular global mode of oscillation corresponds to high-latitude velocity patterns that were reported previously by \citet{ulrich2001} and \citet{hathaway2013}.

Later, \citet{hanson2022} reported the detection of an additional family of inertial modes in the $\ell = m + 1$ component of the radial vorticity spectrum. These modes exhibit north-south antisymmetric vorticity patterns near the equator and propagate in the retrograde direction at frequencies approximately three times higher than those of the equatorial Rossby modes. 
Because their frequencies deviate strongly from those of the classical Rossby modes toward more retrograde direction, they were termed high-frequency retrograde (HFR) modes by \citet{hanson2022}.

A number of observational studies  further characterized the properties of solar inertial modes using a variety of helioseismic techniques and datasets \citep[e.g.,][]{liang2019,hanasoge2019,proxauf2020,hathaway2021,hanson2025, liang2025,lekshmi2026}.
These works have refined measurements of mode frequencies, amplitudes, linewidths, and temporal variability, but they largely confirm the overall observational picture established by the initial detections.

The diversity of observed inertial modes and their departures from classical theoretical expectations immediately highlighted the need for their theoretical and numerical modeling. 
In the remainder of this review, we therefore focus on recent advances in the theoretical modeling of solar inertial modes, which provide the physical framework necessary to understand these observations and to exploit their diagnostic and dynamical potential.
We will show, in particular, that the effects of solar differential rotation, background stratification, entropy gradients, and turbulent diffusion are essential for interpreting the observed mode frequencies and eigenfunctions. 
For an overview of observational aspects of solar inertial modes, we refer the reader to \citet[][Sect.~2]{gizon2024iau} and \citet[][Sect.~9 and references therein]{kosovichev2025}.

To further clarify the scope of this review, we note that the main focus is on the theoretical modeling of \textit{non-magnetic} nature of solar inertial modes. This is because magnetohydrodynamic (MHD) effects have only begun to be explored in a few preliminary studies, and a systematic theoretical framework has not yet been established. 
Furthermore, this review will primarily focus on the modes that are relevant to the recent observations. 
That is, the modes for which observational evidence remains inconclusive, such as thermal Rossby modes \citep[e.g.,][]{glatzmaier1981,busse2002} or wave-attractor-type solutions discussed extensively in the literature \citep[e.g.,][]{rieutord1997,rieutord2001}, are only briefly mentioned in this review.

\section{Definition and Classification of Inertial Modes} \label{classification}

Inertial waves are traveling waves in rotating fluids whose restoring force is the Coriolis force. 
In an unbounded medium, inertial waves form a continuous spectrum of frequencies within the so-called inertial frequency range $\omega \in [-2\Omega_0,\,2\Omega_0]$, where $\Omega_0$ is the rotation rate. 
However, when inertial waves are trapped within a bounded domain such as a sphere or a spherical shell, they can no longer have a continuous spectrum because the boundary conditions select discrete resonant solutions that form global standing oscillations. 
These discrete global modes of oscillation are referred to as \emph{inertial modes}.
In this review, we consistently use the term \textit{waves} to describe local propagation of wave phenomena, and the term \textit{modes} to denote the Sun’s global oscillations that are discretized by boundary conditions and characterized by quantum numbers such as the azimuthal order $m$.

\subsection{Wave-Attractor-Type Solutions}

Inertial modes in a spherical shell geometry have been extensively studied under simplifying assumptions such as incompressible fluid and uniform rotation.
In such a case, the inviscid boundary-value problem becomes ill-posed and singularities exist in solutions due to the hyperbolic nature of the governing equation for general inertial waves, the so-called Poincaré equation \citep{stewartson1969}. 
When a small viscosity is included, these singularities can be regularized into thin internal shear layers aligned with characteristic ray paths determined by the mode frequency. 
These characteristic rays are known to converge toward the so-called attractors, leading to spatially localized eigenfunctions that cannot be well represented by spherical harmonics \citep[e.g.,][]{rieutord1997,tilgner1999,rieutord2001}.
We note that flow motions associated with these inertial modes are essentially three-dimensional (3D).

Prior to the recent discoveries of various inertial modes on the Sun, theoretical studies of inertial modes had largely focused on these wave-attractor-type solutions.
For instance, these studies investigated the effects of background stratification \citep[][]{dintrans1999,dintrans2000}, differential rotation \citep[][]{baruteau2013,guenel2016,mirouh2016}, and tidal forcing \citep[][]{ogilvie2005,rieutord2010}.
In this review, however, these wave-attractor-type inertial modes are not discussed any further because no clear observational counterparts of such modes have so far been identified in solar surface measurements. 
According to linear models of the solar inertial modes (\S~\ref{sec:linear}), these spatially localized modes are more strongly damped than the inertial modes observed on the Sun and discussed in the remainder of this review.
Therefore, the wave-attractor-type modes are expected to contribute less prominently to the low-frequency power spectrum at the surface.

\subsection{Quasi-Toroidal Assumption and Classical Rossby Modes} \label{sec:quasi-toroidal}

Rossby modes belong to a special subclass of inertial modes, namely, a \textit{quasi-toroidal} class of inertial modes in which the associated radial motions are negligible compared to the horizontal ones \citep[][]{rieutord2001b}.
In the case of uniform rotation, the eigenvalue problem of inertial modes in a spherical shell can be significantly simplified with the toroidal approximation, where the governing Poincaré equation is shown to have regular analytical solutions in terms of spherical harmonics \citep[][]{rieutord2001}.
These special solutions correspond to Rossby modes.
In the context of stellar non-radial oscillations, the toroidal class of global oscillations are called \textit{r modes} \citep[][]{papaliozou1978} and interpreted as the stellar counterparts of Rossby waves in the Earth's atmosphere and oceans \citep[e.g.,][]{saio1982,unno1989}. 
Under uniform rotation (with rotation rate $\Omega_0$), their horizontal velocity eigenfunctions are expressed by a toroidal scalar (stream function) proportional to a single spherical harmonic $Y_\ell^m$, and their dispersion relation is given analytically as \citep[][]{provost1981,saio1982}
\begin{equation}
\omega = -\frac{2 m \Omega_0}{\ell(\ell+1)}. \label{eq:rossby}
\end{equation}
Since $m \ge 0$ by convention, negative frequencies $\omega$ denote that they propagate in a retrograde direction (opposite to rotation) in the co-rotating frame.
In this review, we shall refer to the modes described by Eq.~(\ref{eq:rossby}) as \textit{classical} Rossby modes, in order to distinguish them from more generalized forms of Rossby waves discussed in \S~\ref{sec:rossby}.
Moreover, classical Rossby modes are characteristic of uniformly rotating systems and are fundamentally distinct from modes such as critical-layer modes or high-latitude modes, both of which are strongly shaped by differential rotation (\S~\ref{sec:diffrot}).

We note that the existence of this analytical solution played an important role in the initial observational identification of Rossby modes on the Sun \citep[e.g.,][]{loeptien2018}. 
However, as described in \S~\ref{sec:intro}, only a ridge corresponding to the sectoral ($\ell = m$) component was found to follow Eq.~(\ref{eq:rossby}).
According to the semi-analytical work by \citet{damiani2020}, this is because the non-sectoral classical Rossby modes cannot maintain the radial force balance required for their quasi-toroidal nature when the background stratification is close to adiabatic.
Similar conclusions were implied by \citet[][in Appendix B]{rieutord1997}.
The solar observations further showed that, although the mode is strongly localized near the equator, their spatial distributions exhibit substantial non-sectoral contributions in contrast to the classical theory \citep[][]{loeptien2018,proxauf2020}.
This also indicates that the theoretical framework of classical Rossby modes is oversimplified to describe the real Sun, where background stratification, differential rotation, turbulent viscosity, and possibly magnetic fields are expected to play significant roles. 
For this reason, the Rossby modes observed on the Sun are conventionally referred to as \textit{equatorial} Rossby modes \citep[][]{loeptien2018}, rather than the sectoral components of the idealized classical Rossby modes.


\subsection{Various ``Rossby Waves'' Across Disciplines} \label{sec:rossby} 

The term ``Rossby waves'' has been used to describe different physical phenomena across multiple disciplines \citep[e.g.,][]{pedlosky1987,zaqarashvili2021}. 
To avoid terminological confusion, we briefly review its usage in adjacent fields and clarify the definition adopted in this review.

Originally, Rossby waves \citep[][]{rossby1939,rossby1940} were discovered as retrograde-propagating global-scale waves in the Earth's atmosphere and oceans.
In this field, the flow motions are essentially two-dimensional (2D) and they owe their existence to the so-called planetary $\beta$-effect, i.e., the latitudinal gradient of the Coriolis parameter \citep[e.g.,][]{vallis2006}.
In geodynamo and planetary core dynamics, on the other hand, a distinct class of inertial waves are often referred to as \textit{thermal} Rossby waves. 
They are prograde-propagating vortical columns aligned with the rotational axis near the equator, which repeatedly appear near the onset of thermal convection in a rotating spherical shell under the Boussinesq approximation \citep[][]{busse1970,zhang1992,jones2000}. 
Although they involve 3D flow patterns with substantial radial motions, they still belong to a family of Rossby waves because their prograde propagation arises as a consequence of the so-called topographic $\beta$-effect, i.e., the geometrical variation of column height induced by spherical curvature \citep[e.g.,][]{busse2002}.
In stratified systems such as the Sun or gas giants, the background density variation also introduces an additional restoring mechanism for thermal Rossby waves, sometimes referred to as a \textit{compressional} $\beta$-effect \citep[][]{glatzmaier1981,glatzmaier2009,verhoeven2014}. 
All these classes of Rossby waves can be understood within a unified framework based on potential vorticity conservation, which provides the most general definition of Rossby waves \citep[][]{pedlosky1987,miesch2005,zaqarashvili2021}.
Needless to say, all these generalized Rossby waves belong to inertial waves because their restoring force is the Coriolis force.

In models of solar inertial modes, one obtains a rich spectrum of global eigenmodes that includes, but is not limited to, modes with Rossby-type characteristics described above. 
Although $\beta$-effects offer useful physical insights, most inertial modes cannot be uniquely classified within the geophysical taxonomy of generalized Rossby waves. 
As will be shown in \S~\ref{sec:linear}, the properties of solar inertial modes are strongly influenced by internal differential rotation and entropy gradients, making it even more difficult to apply a theoretical taxonomy derived from idealized models.
Because the study of solar inertial modes has been largely observationally driven, the terminology that has emerged reflects measurable properties on the Sun (see \S~\ref{sec:intro}). 
In this review, we follow this pragmatic approach and avoid introducing the broader geophysical classification. 
Instead, we will stick to the more restricted and operational definition of Rossby modes introduced in \S~\ref{sec:quasi-toroidal}.


\section{Linear Models} \label{sec:linear}  

In this section, we discuss linear eigenmode calculations of the Sun's global oscillations in the inertial frequency range.
Linear models are valuable because they yield mode frequencies and spatial eigenfunctions that can be directly compared with observations.
Moreover, they enable systematic investigation of how these observables depend on underlying physical parameters such as the rotation profile, stratification, and unknown transport coefficients, thereby providing a theoretical basis for probing the solar interior.

Linear models of solar inertial modes can be broadly divided into 1.5D reduced models and more complete 2.5D spherical-shell models.
Here, the additional 0.5D denotes the longitudinal dependence expressed in Fourier space.
The 1.5D models restrict the numerical domain to either the $\beta$-plane \citep{gizon2020}, the spherical surface \citep[e.g.,][]{fournier2022}, or an equatorial plane \citep[e.g.,][]{hindman2022,hindman2023}, which allows detailed exploration of selected modes at low computational cost.
A more realistic 2.5D model incorporating differential rotation and background stratification was first presented by \citet{bekki2022a}. 
Subsequent studies have explored similar formulations and extended the parameter space in various directions \citep[e.g.,][]{triana2022,bhattacharya2023,bhattacharya2024,mukhopadhyay2025,mukhopadhyay2026}.
These linear models provide the essential baseline for interpreting the observed solar inertial modes.

\subsection{Linear Eigenvalue Problem: Model Equations} \label{sec:eigenvalue}

In this section, we briefly outline the 2.5D formulation largely following \citet{bekki2022a}.
Throughout, we adopt standard notation in solar physics unless otherwise specified \citep[see][for a full specification of the model]{bekki2022a}.
The governing equations consist of the linearized hydrodynamic equations -- namely, equations of continuity, motion, and entropy -- formulated in the frame co-rotating with the angular velocity $\Omega_0$ and written in spherical coordinates $(r,\theta,\phi)$ as 
\begin{eqnarray}
    && \frac{\partial \rho'}{\partial t} = -\nabla\cdot(\rho\bm{v}')-\nabla\cdot(\rho'\bm{v}), \label{eq:continuity} \\
    && \frac{\partial \bm{v}'}{\partial t} = -\bm{v}\cdot\nabla\bm{v}' - \bm{v}'\cdot\nabla\bm{v} -\frac{\nabla p'}{\rho}-\frac{\rho'}{\rho}g\bm{e}_r - 2\Omega_0\bm{e}_z \times \bm{v}' + \frac{1}{\rho}\nabla\cdot (\nu_t \rho \mathbf{S}), \label{eq:motion} \\
    && \frac{\partial s'}{\partial t} = -\bm{v}\cdot\nabla s' -v'_r \frac{\partial s}{\partial r} - \frac{v'_\theta}{r}\frac{\partial s}{\partial \theta} + \frac{1}{\rho T} \nabla\cdot (\kappa_t \rho T  \nabla s'). \label{eq:entropy}
\end{eqnarray}
Here, $\bm{v}'$, $\rho'$, $p'$, and $s'$ are the perturbations of velocity, density, pressure, and entropy with respect to the time-independent background (without primes). 
While the background stratification can be specified according to the solar standard model \citep[e.g.,][]{christensen1996}, the radial entropy gradient $\partial s/ \partial r$, representing the degree of convective instability in the Sun's convection zone, remains largely unknown. 
The time-independent background mean flow $\bm{v}=\bm{v}_{\rm m} + (\Omega - \Omega_0)r\sin{\theta} \bm{e}_\phi$ represents the meridional flow and differential rotation.
The meridional flow has been shown to have negligible impact on most inertial modes \citep[][]{gizon2020,fournier2022} and thus often omitted for simplicity.
The profile of differential rotation $\Omega(r,\theta)$ is well constrained by global helioseismology \citep[e.g.,][]{larson2018} except near the poles where the inversion with p modes becomes less reliable \citep[][]{dey2026a}.
The latitudinal entropy gradient is typically specified under the assumption of thermal wind balance \citep[e.g.,][]{thompson2003,miesch2006},
\begin{eqnarray}
    && \frac{g}{c_{\rm p}}\frac{\partial s}{\partial\theta} \approx r^2 \sin{\theta} \frac{\partial (\Omega^2)}{\partial z}, \label{eq:twb}
\end{eqnarray}
where $z$ denotes the rotation axis.
Although this balance is well maintained in the bulk of the convection zone, it is likely not satisfied in the near-surface shear layer \citep{hotta2025}.
The model also takes into account turbulent diffusion of momentum and entropy, represented by $\nu_{\rm t}$ and $\kappa_{\rm t}$, with the viscous stress written in terms of the traceless strain-rate tensor $\mathbf{S}$.
Although values of turbulent diffusivities are estimated to be on the order of $10^{12}$~cm$^2$~s$^{-1}$ in the Sun's convection zone \citep[e.g.,][]{munoz2011}, their exact profiles remain uncertain.
Finally, we note that the viscous heating term is omitted from the energy equation in most studies for simplicity.

Assuming that all the perturbations are proportional to $\mathrm{exp}[i(m\phi-\omega t)]$, the above linearized Eqs.~\eqref{eq:continuity}--\eqref{eq:entropy} as well as the linearized equation of state are combined into an eigenvalue problem,
\begin{eqnarray}
    \omega \bm{x} = \mathcal{L} \bm{x}, \ \ \  \mathrm{with} \ \ \bm{x}=(\rho'\ v_r'\ v_\theta'\ v_\phi'\ s')^T. \label{eq:eigen}
\end{eqnarray}
Here, $\mathcal{L}$ represents the linear differential operator which depends on azimuthal order $m$, differential rotation profile $\Omega$, radial and latitudinal entropy gradients $\partial s/\partial r$ and $\partial s / \partial\theta$, turbulent diffusivities $\nu_{\rm t}$ and $\kappa_{\rm t}$, and other model parameters.
In practice, the discretized operator reduces to a complex matrix.
This eigenvalue problem is completed by imposing appropriate boundary conditions.
The numerical domain is usually restricted to the bulk of the convection zone $r \in [0.71R_{\odot},0.985R_{\odot}]$ or with the upper half of the radiative interior included.
We note that the near-surface layer is typically excluded because its steep stratification requires exceedingly high resolution, and its strong superadiabaticity renders the problem numerically challenging (see \S~\ref{sec:domain}). 
At both radial boundaries, impenetrable and stress-free boundary conditions with vanishing entropy flux are commonly used.
\citet{mukhopadhyay2026} also verified that adopting free-surface boundary conditions does not significantly alter the results.
At the poles, all the perturbation variables are set to be zero for $m>1$. At $m=1$, the special boundary condition is imposed to explicitly allow for the pole-crossing flow.
Under these boundary conditions, the eigenvalue equation \eqref{eq:eigen} can be solved numerically using either finite-difference method \citep[][]{bekki2022a} or spectral method \citep[][]{triana2022,bhattacharya2023,mukhopadhyay2025}.

We note that the above formulation is not unique among 2.5D models of solar inertial modes. 
While \citet{bekki2022a} adopt a fully compressible formulation in which high-frequency p modes are also retained, \citet{bhattacharya2023} employ the anelastic approximation, and \citet{triana2022} further use an incompressible model to investigate solar inertial modes.
\citet{mukhopadhyay2025} systematically examined the impact of these approximations on solar inertial modes, and demonstrated that the anelastic approximation substantially reduces computational cost without compromising accuracy, whereas the incompressible (Boussinesq) approximation fails to capture essential features of solar inertial modes.
Under the anelastic approximation, \citet{bhattacharya2023} solved the linearized vorticity equation using a spectral method \citep[see, e.g.,][]{rieutord1997}.
In idealized settings, this approach substantially reduces numerical cost and enables efficient computation of large-scale Rossby modes. However, this approach faces numerical difficulties when incorporating realistic effects such as the Sun's differential rotation, the latitudinal entropy gradient, and the sharply varying superadiabaticity \citep[][]{bhattacharya2024}.
Using the flexible open-source code \texttt{Dedalus}, \citet{mukhopadhyay2025,mukhopadhyay2026} took a hybrid approach wherein the linearized equations of motion and entropy are solved in their primitive form using the spectral method.


\subsection{Inertial Modes Under Uniform Rotation} \label{sec:uniformrotation}

\begin{figure}[t]
\centering
\includegraphics[width=\textwidth]{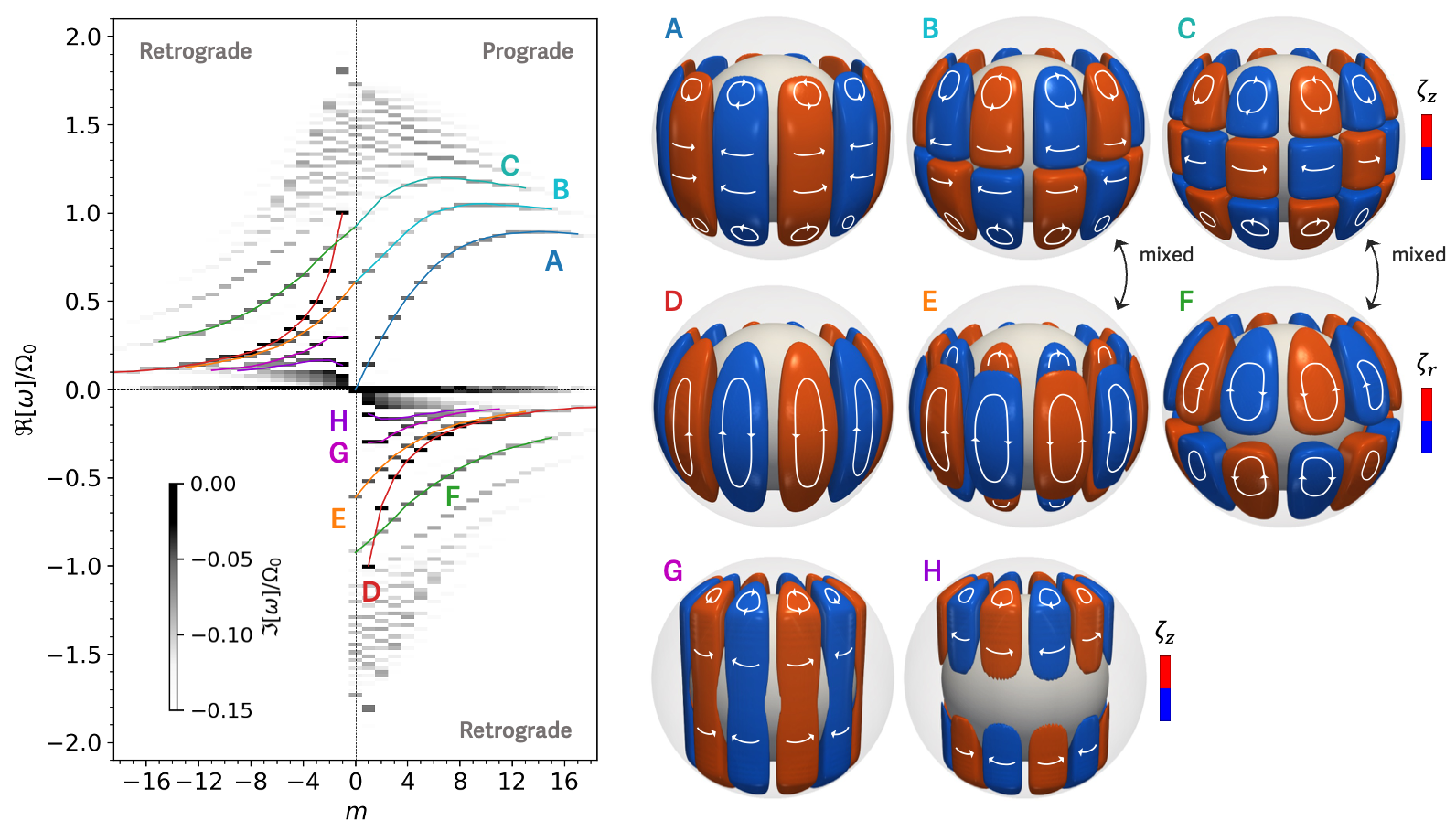}
\caption{
Inertial modes in a uniformly rotating convection zone computed using the 2.5D linear model of \citet{bekki2022a}.
(\textit{Left}) Synthetic spectrum in the inertial frequency range showing frequency-binned mode growth rates, $\Im[\omega]$, in the $m$--$\Re[\omega]$ plane, where negative values of $\Im[\omega]$ indicate damped modes.
To emphasize the dominant branches of the least damped modes, a relatively large viscosity $\nu_{\rm t}=3\times 10^{12}$~cm$^2$~s$^{-1}$ is adopted.
To better visualize mixed modes between prograde and retrograde inertial modes, the dispersion diagram is extended to negative azimuthal orders ($m<0$ with $\Re[\omega]>0$).
(\textit{Right}) Schematic flow structures of representative modes at $m=6$.
The modes labeled A--H correspond to:
(A) a prograde columnar mode in which the axial vorticity $\zeta_z$ has no latitudinal nodes;
(B,C) similar prograde columnar modes with one and two latitudinal nodes, respectively;
(D) a sectoral Rossby mode in which the radial vorticity $\zeta_r$ has neither latitudinal nor radial nodes;
(E) similar to D but with one radial node;
(F) similar to D but with one latitudinal node;
(G,H) modes strongly confined inside the tangent cylinder, in which the axial vorticity $\zeta_z$ is symmetric and antisymmetric across the equator, respectively.
Red and blue surfaces indicate isocontours of $\zeta_z$ in panels A--C and G--H, and of $\zeta_r$ in panels D--F.
The modes corresponding to C and F were not discussed in \citet{bekki2022a}.
}\label{fig:uniformrotation}
\end{figure}

Before discussing the realistic models including the Sun's differential rotation and internal entropy gradients, it is instructive to briefly overview inertial modes under uniform rotation ($\Omega=\Omega_0$) with adiabatic background ($\partial s/\partial r = \partial s / \partial \theta =0$).
As has been shown by \citet{rieutord2018}, the spectrum of inertial modes is remarkably rich even in this idealized case. 
Part of this complexity arises because spherical-shell geometry permits not only equatorially trapped modes but also modes confined within the tangent cylinder.
Furthermore, each family of modes spans multiple radial and latitudinal overtones.

Figure~\ref{fig:uniformrotation} (left panel) exemplifies the synthetic spectrum in the inertial frequency range where the frequency-binned damping rates are mapped in the $m-\omega$ space.
This representation effectively highlights the dominant branches of the least damped modes of the system.
In what follows, we summarize the main classes of modes that are relevant for solar applications.

\subsubsection*{-- Equatorial Rossby Modes}

Equatorial Rossby modes are retrograde-propagating modes of radial vorticity $\zeta_r$ concentrated near the equator.
They can be identified as the least damped class of inertial modes (labeled as D in Fig.~\ref{fig:uniformrotation}), following the analytical dispersion relation, $\Re[\omega]/\Omega_0 = -2/(m+1)$.
Under uniform rotation, they correspond to the sectoral components of the classical Rossby modes.
In the adiabatically stratified convection zone, only the equatorial Rossby modes without radial nodes ($n=0$) remain purely toroidal, because other non-sectoral modes or modes with radial nodes fail to satisfy the radial force balance \citep[e.g.,][]{damiani2020}.
We note that the $m=1$ mode has the same retrograde frequency as the solar rotation rate ($\Re[\omega]=-\Omega_0$) with uniform radial vorticity directing perpendicular to the rotational axis. 
This particular mode is often called spin-over mode \citep[e.g.,][]{greenspan1968}.

\citet{bekki2022a} identified another class of modes that asymptotically approach the $n=0$ equatorial Rossby modes in the limit of large $m$.
They are labeled as E in Fig.~\ref{fig:uniformrotation}. 
These modes possess equatorially symmetric radial vorticity concentrated at low latitudes, similar to the $n=0$ equatorial Rossby modes, but are characterized by the presence of a radial node ($n=1$) in the radial vorticity $\zeta_r$.
The existence of the radial node inevitably induces weak radial motions, leading to a deviation from strictly toroidal mode.
We conventionally refer to them as $n=1$ equatorial Rossby-like modes.
It should be noted that the associated nodal surface does not extend in latitude but is instead aligned approximately parallel to the rotation axis.
For small $m$, these modes propagate more slowly in the retrograde direction than the $n=0$ modes. 
However, for $m \gtrsim 8$, their frequencies become nearly indistinguishable from those of the $n=0$ modes.
As will be shown later, these $n=1$ modes are essentially mixed with prograde columnar modes with one latitudinal node.

\subsubsection*{-- High-Frequency Retrograde (HFR) Modes}

Under uniform rotation, linear models predict a variety of retrograde-propagating modes of radial vorticity. 
These modes propagate more rapidly in the retrograde direction than the equatorial Rossby modes and their frequencies increase with the number of latitudinal nodes $\lambda$, clearly indicating that they are fundamentally distinct from the non-sectoral classical Rossby modes that are purely toroidal.
The modes corresponding to the HFR modes detected by \citet{hanson2022} were identified as one such branch (labeled F in Fig.~\ref{fig:uniformrotation}), whose radial vorticity eigenfunction has one latitudinal node ($\lambda=1$) across the equator \citep[][]{triana2022,bhattacharya2023}.
The HFR modes and other $\lambda \ge 1$ modes turn out to be strongly non-toroidal, i.e., substantial radial motions are involved as well as radially vortical motions.
Observational evidence for such non-toroidal nature of the HFR modes has also been reported by \citet{hanson2025}.
To understand why the HFR modes propagate in the retrograde direction faster than the classical Rossby modes, it is instructive to examine the radial vorticity equation \citep[][]{mukhopadhyay2025}
\begin{eqnarray}
    \frac{\partial \zeta_r}{\partial t} = \frac{2\Omega_0 \sin{\theta}}{r} v_\theta -\frac{2\Omega_0 \cos{\theta}}{H_{\rho}}v_r  +[\ ... \ ], \label{eq:radvort}
\end{eqnarray}
where $H_{\rho}$ is the density scale height. 
The first term on the right-hand side represents the planetary $\beta$-effect which results in the retrograde mode propagation similarly to the Rossby modes.
For non-toroidal modes, however, the second term on the right-hand side plays a significant role.
Since the HFR modes have upflows ($v_r > 0$) where latitudinal velocity diverges ($\partial v_\theta/ \partial\theta >0$) near the equator, the second term leads to a generation of radial vorticity perturbation which is negative (positive), $\partial \zeta_r / \partial t <0 \ (>0)$, in the upflow regions in the northern (southern) hemisphere. 
The opposite happens in downflow regions with latitudinal converging flow.
Consequently, the second term enhances the overall retrograde propagation of the HFR modes.

\subsubsection*{-- Prograde Columnar Modes}

Linear models predict several classes of prograde-propagating inertial modes, labeled A–C in Fig.~\ref{fig:uniformrotation}.
These modes have dominant flow motions characterized by columnar vortex motions outside the tangent cylinder aligned with the rotational axis ($z$-direction).
Among the prograde modes, the least-damped modes are those for which the axial vorticity $\zeta_z$ has no nodes in latitude, $\lambda=0$ (labeled A in Fig.~\ref{fig:uniformrotation}), where $\lambda$ denotes the number of latitudinal nodes.
Near the equator, strong radial motions are involved where longitudinal flow converges or diverges.
The corresponding flow structure is reminiscent of the well-known Busse columns \citep[e.g.,][]{busse1970} seen in Boussinesq models of rotating convection in a spherical shell.
These Busse columns are also referred to as \textit{thermal} Rossby waves, as they appear at the onset of thermally driven convection and their propagation arises from the topographic $\beta$-effect \citep[e.g.,][]{busse2002}.
However, in the Sun's convection zone, the dynamics are dominated by the compressional $\beta$-effect arising from background density stratification \citep[][]{glatzmaier1981,bekki2022a,hindman2022}, which fundamentally distinguishes these modes from their incompressible counterparts. Moreover, since the present modes are obtained even under adiabatic stratification, their existence does not rely on thermal buoyancy. 
The qualifier ``thermal'' might therefore be misleading when describing the general properties of the prograde columnar modes.

Prograde columnar modes can have structures in latitude (or in $z$).
The modes with one and two latitudinal nodes ($\lambda=1,2$) are shown in Fig.~\ref{fig:uniformrotation} (labeled B and C, respectively).
In general, modes with a larger $\lambda$ exhibit higher prograde frequencies, although their frequencies asymptotically converge to those of the fundamental ($\lambda=0$) branch in the limit of large $m$.
The $\lambda=1$ modes with axial vorticity antisymmetric across the equatorial plane were also explored in early Boussinesq studies as convective onset modes \citep[e.g.,][]{lorenzani2001}.
Later, \citet{bekki2022a} pointed out that these prograde columnar modes with $\lambda=1$ are mixed with the $n=1$ equatorial Rossby-like modes across $m=0$.

\subsubsection*{-- Mixed Inertial Modes}

Figure~\ref{fig:uniformrotation} (left) clearly shows that many prograde and retrograde inertial-mode branches are connected across $m=0$, forming continuous dispersion branches that can be interpreted as \textit{mixed} inertial modes.
Such mode mixing can be understood in analogy with the so-called Yanai waves seen in the Earth's equatorial atmosphere, which are mixed modes between retrograde Rossby modes and prograde inertial-gravity modes \citep[][]{matsuno1966,vallis2006}.
\citet{bekki2022a} were the first to point out the existence of these mixed modes in the spherical shell model of the Sun, who found that the prograde columnar modes with one latitudinal node ($\lambda=1$) are mixed with the retrograde-propagating equatorial Rossby-like modes with one radial node ($n=1$).
\citet{jain2023,jain2024} further proposed a generalized picture of mixed inertial modes with larger latitudinal overtones, building on the 1.5D cylindrical model of \citet{hindman2022} while incorporating vertical structure via Fourier decomposition.
Within this picture, the HFR modes also correspond to the retrograde branch of the mixed modes with 2nd latitudinal overtones ($\lambda=2$).

How can oppositely-propagating prograde and retrograde modes be mixed with each other? 
Insight can be gained from the 3D eigenfunctions of the $\lambda=1$ mixed modes \citep[e.g., Figs.~13 and 14 of][]{bekki2022b}. 
These figures show that the two branches share nearly identical radial and latitudinal velocity structures, while differing in the sign of the longitudinal velocity.
In the axisymmetric case ($m=0$) where the modes are non-propagating, the two branches become degenerate and correspond to complex-conjugate representations of the same eigenfunction. 
From a physical perspective, both branches arise from a common eigenfunction of the pressure perturbation $p'$ characterized by the same number of radial and latitudinal nodes \citep[$n=1$ and $\lambda=1$ in the case shown in][]{bekki2022a,bekki2022b}.
At $m \neq 0$, this pressure structure can support two distinct quasi-geostrophic balances depending on the orientation of the velocity field. 
In one case, the pressure gradient force is primarily balanced by the Coriolis force associated with $z$-vortical motions, leading to the prograde columnar mode largely dominated by the compressional $\beta$-effect. 
In the other, the balance involves a different projection of the Coriolis force linked to the radial vorticity, leading to the $n=1$ equatorial Rossby-like branch.
Similar explanation was also presented by \citet{ong2025}.
The origin of the mixed inertial modes can also be understood in terms of the wave topology \citep{leclerc2024}.

\subsubsection*{-- High-Latitude Modes}

So far, we have mainly focused on modes near the equatorial region (A--F).
Now, we discuss modes that reside inside the tangent cylinder.
The modes labeled G and H in Fig.~\ref{fig:uniformrotation} were identified as columnar modes characterized by axial vorticity $\zeta_z$ extending along the rotation axis within the tangent cylinder \citep[][]{bekki2022a}. 
Modes G and H have north-south symmetric and antisymmetric $\zeta_z$, respectively.
Since these modes are observed near the poles at the surface, we refer to them as \textit{high-latitude} modes.
They propagate slowly in the retrograde direction under uniform rotation. 
This retrograde propagation is likely due to the topographic $\beta$-effect arising from the curvature of the inner spherical boundary. 
Their frequencies decrease as the number of latitudinal nodes $\lambda$ increases.

It is important to note that the high-latitude modes are strongly affected by the Sun's differential rotation. 
As will be discussed later, modes resembling the high-latitude modes are sometimes identified in 1.5D spherical surface models; however, those toroidal modes are physically distinct. 
The high-latitude modes obtained in the 2.5D spherical shell model involve an intrinsically 3D structure.


\subsection{Effects of Differential Rotation} \label{sec:diffrot}

\begin{figure}[t]
\centering
\includegraphics[width=\textwidth]{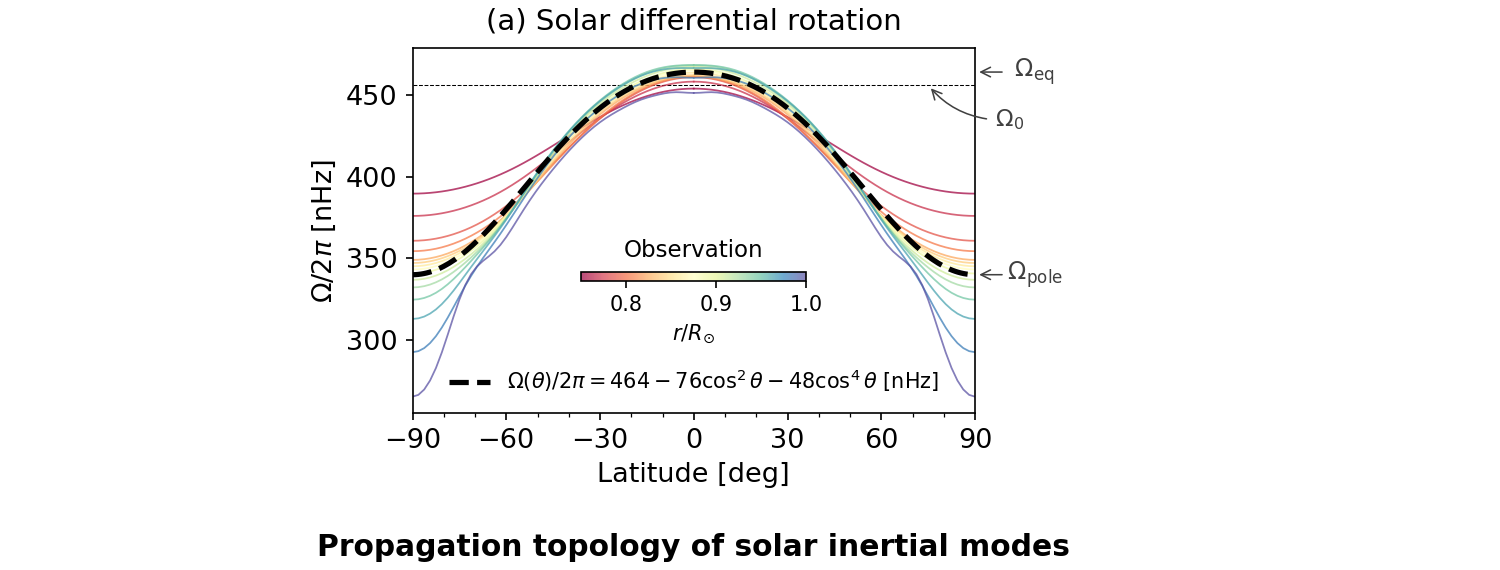}
\includegraphics[width=\textwidth]{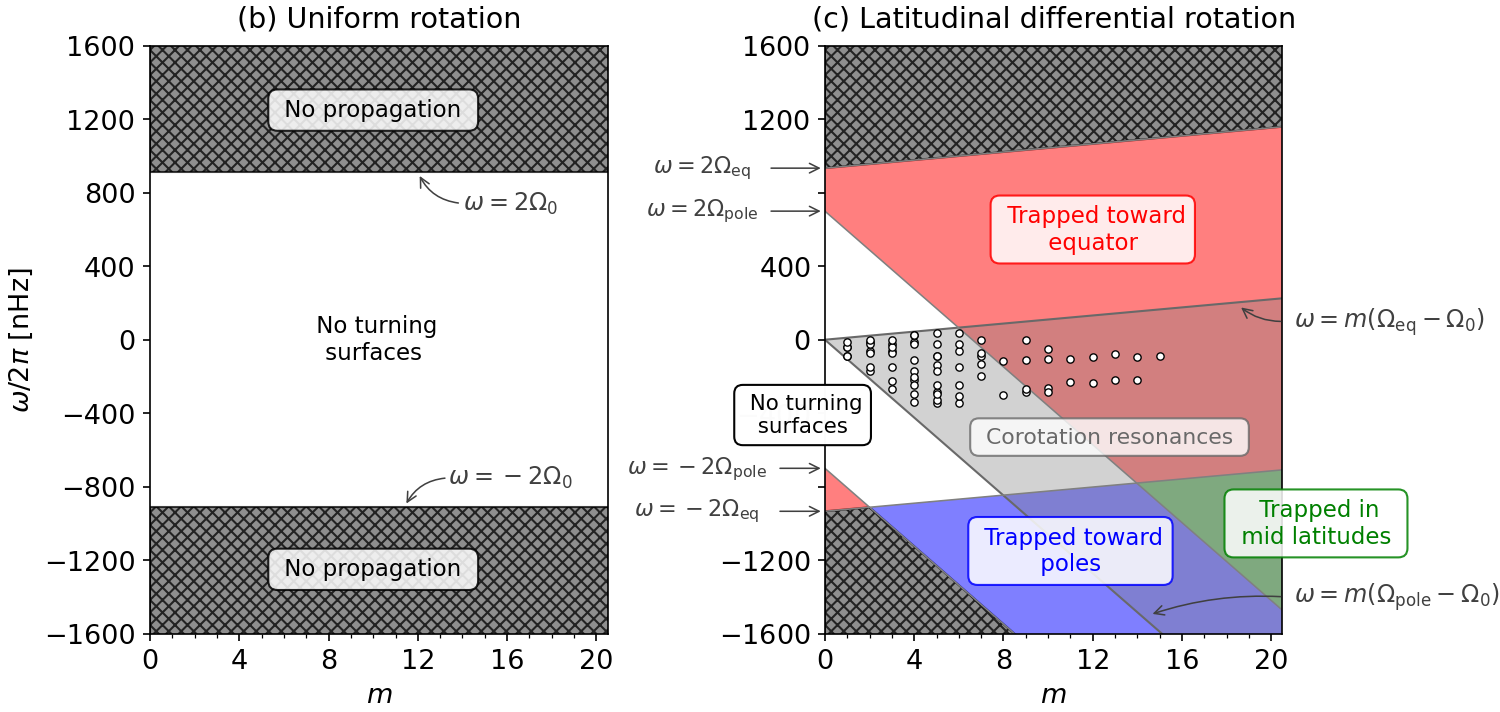}
\caption{
Impact of the Sun's differential rotation on propagation topology of inertial modes.
(a) Latitudinal profiles of the Sun's differential rotation at different depths \citep[][]{larson2018}. 
Black solid curve denotes the fitted analytical profile.
(b) Propagation domain of inertial modes under uniform rotation in the $(m,\omega)$ plane. Frequencies are measured in a Carrington frame corotating at $\Omega_0/2\pi=456$~nHz.
No inertial modes exist at $|\omega| > 2\Omega_0$.
(c) Same as panel b, but in the case with the latitudinal differential rotation.
The differently colored regions indicate different propagation topologies:
\textit{White}; modes without turning surfaces.
\textit{Red}; modes trapped toward the equator.
\textit{Blue}; modes trapped toward the poles.
\textit{Green}; modes trapped at mid-latitudes by turning surfaces.
The gray shaded area indicates the frequency range in which modes can be affected by viscous critical layers associated with corotation resonances.
The white circles show the observed frequencies of the equatorial Rossby, HFR, high-latitude, and critical-layer modes (see Fig.~\ref{fig:gizon21_comp} for details).
}\label{fig:turning_surfaces}
\end{figure}

\begin{figure}[t]
\centering
\includegraphics[width=0.9\textwidth]{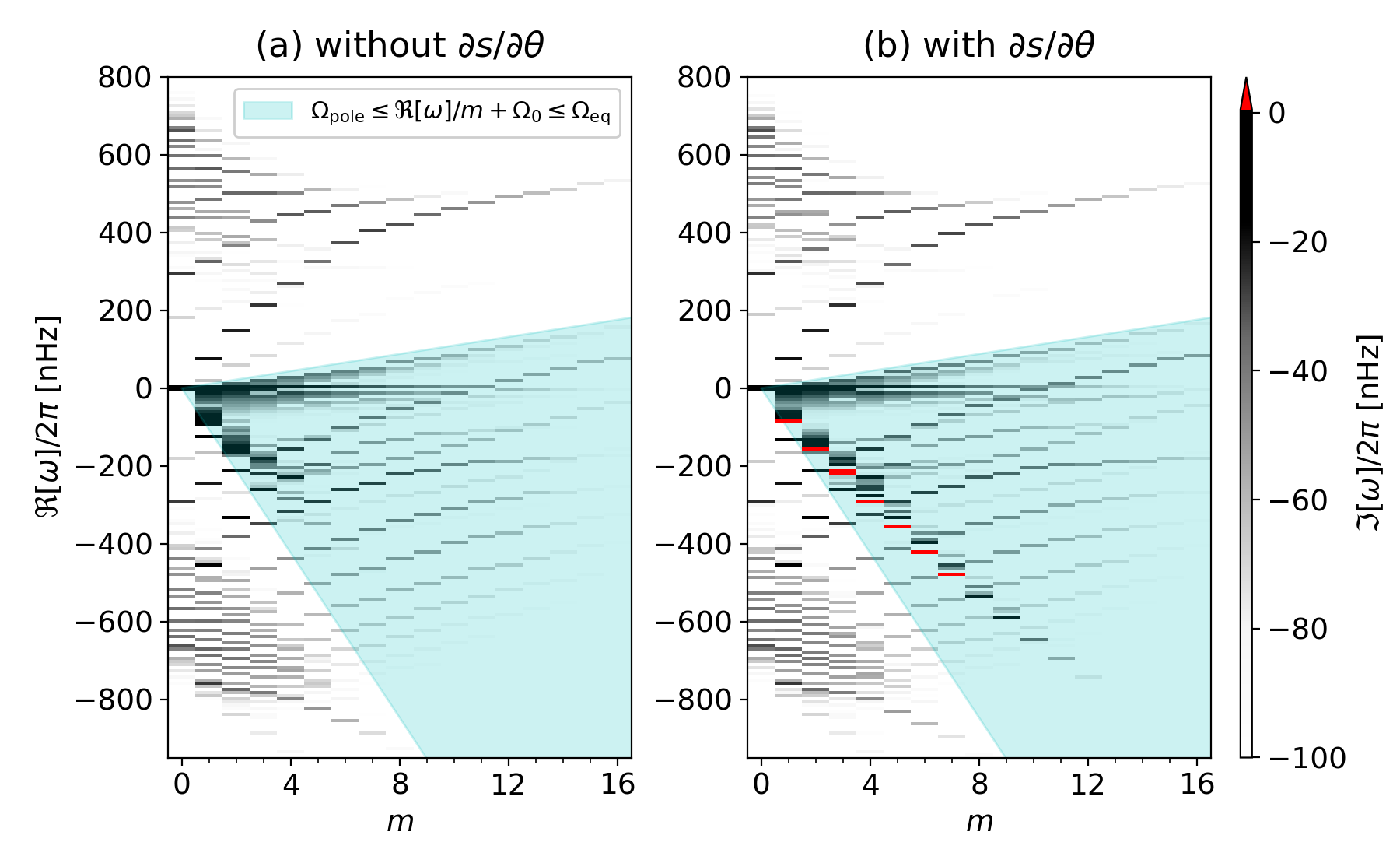}
\caption{
Same synthetic spectrum of inertial modes from the 2.5D model of \citet{bekki2022a} as Fig.~\ref{fig:uniformrotation} but under solar differential rotation \citep[from][]{larson2018}.
The frequencies are measured in the Carrington frame ($\Omega_0 / 2\pi = 456$~nHz).
The same viscous and thermal diffusivities ($\nu_{\rm t}=\kappa_{\rm t}=3.0\times 10^{12}$~cm$^2$~s$^{-1}$) are used, corresponding to the Ekman number $\mathrm{Ek}=\nu_{\rm t}/(R_{\odot}^2 \Omega_{0})=2.2\times 10^{-4}$.
The background stratification is adiabatic ($\partial s/\partial r = 0$) for simplicity.
For reference, the cyan-shaded areas indicate the frequency ranges in which modes are influenced by corotation resonances induced by differential rotation, $\Omega_{\rm pole}-\Omega_0 \leq \Re[\omega]/m \leq \Omega_{\rm eq}-\Omega_0$, where $\Omega_{\rm eq}/2\pi = 467$~nHz and $\Omega_{\rm pole}/2\pi = 350$~nHz.
Panels a and b correspond to the cases without and with the latitudinal entropy gradient, $\partial s / \partial \theta$, estimated from Eq.~\eqref{eq:twb}.
Linearly unstable modes (with $\Im[\omega]>0$) are visualized by red colors. 
}\label{fig:diffrot}
\end{figure}

Differential rotation causes a Doppler frequency shift for inertial modes, as it does for p modes, but in addition it also leads to fundamental modifications in their spectrum and spatial structure.
The effects of differential rotation on inertial modes in spherical shells have been investigated in detail by \citet[][]{baruteau2013} and \citet{guenel2016}. 
In the presence of differential rotation
$\Delta\Omega(r,\theta)=\Omega(r,\theta)-\Omega_0$, inertial modes can be affected by corotation resonances and turning surfaces, depending on their frequencies.
Turning surfaces separate the domains where the governing equation is hyperbolic from those where it is elliptic, thereby restricting the regions in which inertial waves can propagate.
Since the Sun's differential rotation is known to be predominantly latitudinal and relatively weak ($|\Delta\Omega|/\Omega_0 \sim 0.25$), as shown in Fig.~\ref{fig:turning_surfaces}a, the condition for local inertial-wave propagation can be approximated (within an error of a few percent) as $|\omega-m\Delta\Omega(r,\theta)| < 2\Omega(r,\theta)$ \citep[see][for more details]{baruteau2013}.
Figures~\ref{fig:turning_surfaces}b and \ref{fig:turning_surfaces}c illustrate how the Sun's latitudinal differential rotation modifies the propagation topology of inertial modes.
We note that most solar inertial modes observed to date fall in the regime where turning surfaces do not play a significant role.
Instead, their mode structures are much more strongly affected by viscous critical layers associated with corotation resonances.
This is also consistent with the results of \citet{guenel2016}, who found modes trapped by turning surfaces to be less common for solar-like differential rotation.
Some prograde columnar modes may nevertheless possess turning surfaces, but these modes have not yet been unambiguously observed in the Sun (see \S~\ref{sec:open}).
Therefore, we do not discuss the effects of turning surfaces further in this review.

A corotation resonance occurs where the mode phase speed matches the local differential rotation speed, $\omega - m \Delta\Omega(r,\theta) =0$, in the rotating frame \citep[e.g.,][]{baruteau2013,guenel2016}. 
This corresponds to locations where the total derivative, $\partial_t + \Delta\Omega \partial_\phi$, vanishes.
In the absence of viscous diffusion, the eigenvalue problem of inertial modes becomes singular along the corotation surfaces \citep[e.g.,][]{balmforth1995,gizon2020}.
The inclusion of viscosity regularizes this singularity and leads to the formation of the so-called viscous critical layers.
Figure~\ref{fig:diffrot}a shows the synthetic inertial mode spectrum with the presence of the Sun's differential rotation.
It is clearly seen that many inertial modes are strongly influenced by differential rotation via viscous critical layers, particularly retrograde modes at high $m$ (as indicated by the cyan-shaded area).
In the vicinity of the critical layers, the mode structures become significantly distorted.
For example, equatorial Rossby modes, which exist as purely sectoral modes under uniform rotation, become increasingly confined toward the equator by the critical layers at large $m$, leading to a pronounced departure from sectoral structure \citep[][]{gizon2020,fournier2022}.
We note that, for these equatorial Rossby modes, the turning surfaces are always located at higher latitudes than the critical layers and therefore have little effect on the mode structure.

\begin{figure}[t]
\centering
\includegraphics[width=0.65\textwidth]{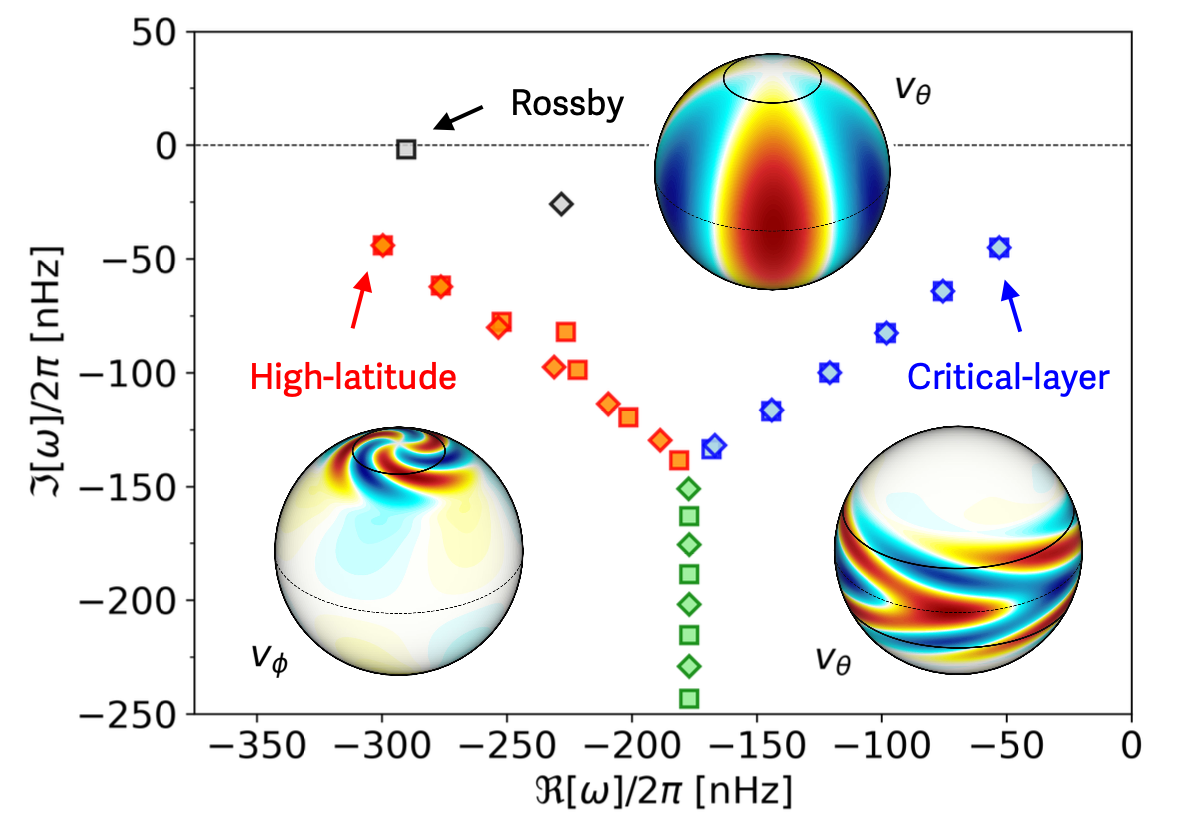}
\caption{
Complex eigenfrequency spectrum of toroidal inertial modes on a spherical surface at $m=3$ computed from the 1.5D  model of \citet{fournier2022} under the latitudinal differential rotation, $\Delta\Omega(\theta) = -\Omega_2 \cos^2{\theta}$ with $\Omega_2 / 2\pi=114$~nHz.
The model includes the turbulent viscosity of $5.5\times 10^{12}$~cm$^2$~s$^{-1}$, corresponding to the Ekman number $\mathrm{Ek}=4\times 10^{-4}$.
The diamonds and square symbols denote modes with north-south symmetric and antisymmetric radial vorticity, respectively.
Gray, red, blue, and green colors denote Rossby, high-latitude, critical-layer, and strongly damped modes.
Example velocity eigenfunctions are also displayed for three selected modes; $v_\theta$ for the equatorial Rossby and critical-layer modes, and $v_\phi$ for the high-latitude mode.
The colatitudes of the critical layers $\theta_c$ are shown by black solid lines.
}\label{fig:critical-latitude}
\end{figure}

The Sun's differential rotation not only modifies the properties of inertial modes present under uniform rotation, but also leads to the appearance of additional branches. 
To isolate the underlying physics, it is instructive to consider simplified 1.5D models in which the dynamics are restricted to toroidal motions.
Using a simplified latitudinal differential rotation profile, $\Delta\Omega(\theta) \propto -\cos^2\theta$, \citet{fournier2022} showed that the eigenfrequency spectrum of toroidal inertial modes exhibits a characteristic Y-shaped structure (Fig.~\ref{fig:critical-latitude}).
In addition to the Rossby modes, three additional branches emerge, corresponding to critical-layer modes, high-latitude modes, and strongly damped modes, all being localized near viscous critical layers.
Here, the value of turbulent viscosity $\nu_{\rm t}$ plays a critical role, as the latitudinal thickness of this layer is shown to scale with the turbulent viscosity as $\propto \nu_{\rm t}^{1/3}$.
With moderate viscosity included, the equatorial Rossby modes exhibit significant Reynolds stresses near the viscous critical layers, leading to equatorward angular momentum transport \citep[][]{gizon2020}.

The high-latitude modes are found to be particularly sensitive to the differential rotation profile near the poles. 
When the actual observed profile of the surface differential rotation is adopted in the 1.5D model (rather than a simplified profile), the high-latitude modes become unstable to shear instability at low $m$ for moderate turbulent viscosity ($\lesssim 10^{13}$~cm$^2$~s$^{-1}$) \citep[][]{fournier2022}. 
However, the resulting high-latitude modes exhibit spiral flow patterns that differ from those observed on the Sun.
This discrepancy arises because the shear-unstable modes have the associated Reynolds stress that transports angular momentum poleward.

Although the 1.5D model offers valuable physical insight, it is too idealized to capture the essential physics of the high-latitude modes. It is therefore important to distinguish between the high-latitude modes identified in 1.5D and 2.5D frameworks.
In reduced 1.5D surface models, the high-latitude branches arise solely from differential rotation through the formation of viscous critical layers. By contrast, in 2.5D spherical-shell models, high-latitude modes already exist as modes trapped inside the tangent cylinder even under uniform rotation (\S~\ref{sec:uniformrotation}).
Differential rotation significantly alters their mode properties. 
This difference underscores the intrinsically 3D nature of the high-latitude modes in spherical shells.
The strong sensitivity of the high-latitude modes to differential rotation also makes them promising probes of the solar internal rotation. 
\citet{dey2026a} used the $m=1$ high-latitude mode to constrain the rotation rate near the poles, which is poorly determined by conventional p-mode helioseismology.
They suggest that the rotation rate at high latitudes ($\sim 70^\circ$) in the middle convection zone ($r\sim0.8R_{\odot}$) is about $8$~nHz higher than inferred from p-mode helioseismology, demonstrating the potential of inertial-mode helioseismology (see \S~\ref{sec:i-helio}).

\subsection{Effects of Baroclinicity} \label{sec:baroclinic}

The solar differential rotation is known to deviate from the Taylor--Proudman constraint. 
This departure is commonly attributed to thermal-wind balance sustained by a small latitudinal temperature contrast within the convection zone \citep[e.g.,][]{thompson2003}, implying that the solar convection zone is intrinsically baroclinic. 
In the 2.5D model, the high-latitude modes are preferentially distributed in regions where the differential rotation exhibits significant radial shear, making them particularly sensitive to such baroclinic effects.

Using turbulent viscosities typical of the solar interior ($\sim 10^{12}$~cm$^2$~s$^{-1}$), \citet{bekki2022a} incorporated the latitudinal entropy gradient inferred from thermal-wind balance (Eq.~\ref{eq:twb}) in addition to the observed differential rotation, and demonstrated that the high-latitude modes are baroclinically unstable at low $m$ \citep[see Fig.~28 in][]{bekki2022a}. 
This is also evident in Fig.~\ref{fig:diffrot}b, where baroclinically unstable modes are highlighted in red in the dispersion diagram.
This linear instability is consistent with the observational fact that these low-$m$ high-latitude modes have the largest amplitudes among the detected inertial modes \citep[][]{gizon2021}. 
Furthermore, unlike the shear-unstable modes in 1.5D models (\S~\ref{sec:diffrot}), the baroclinically unstable modes reproduce the near-surface spiral flow structures which agree well with the observations. 
This difference arises because the baroclinic instability is characterized primarily by equatorward heat transport rather than by poleward angular momentum transport.

\subsection{Effects of Superadiabaticity} \label{sec:delta}

Superadiabaticity $\delta$ is a dimensionless quantity which measures how much the background temperature gradient deviates from the adiabatic one, and is related to the background radial entropy gradient $\partial s/ \partial r$ and the Brunt-Väisälä  frequency $N$,
\begin{eqnarray}
    \delta = -\frac{H_p}{c_p} \frac{\partial s}{\partial r} = -\frac{H_p}{g}N^2,
\end{eqnarray}
where $H_p$ is the pressure scale height and $c_p$ is the specific heat at constant pressure.
Although $\delta$ is well constrained near the photosphere, its profile in the deep convection zone remains largely uncertain.
Recent theoretical studies revealed that some inertial modes are highly sensitive to $\delta$ in the deep solar interior \citep[][]{gizon2021,bekki2022a,bekki2024a,hanson2024}, suggesting that inertial modes may enable observational constraints to be placed on $\delta$.

Prograde columnar modes are known to appear as convective onset modes in rotating convection under superadiabatic background. 
These convectively unstable modes are commonly referred to as thermal Rossby modes \citep{busse1986b}. 
The influence of the superadiabaticity $\delta$ on prograde columnar modes was investigated by \citet{bekki2022a} in detail, where $\delta$ was treated as a spatially constant parameter. 
It was shown that when the convection zone is weakly subadiabatic ($\delta < 0$), the mode frequencies increase (more prograde) for all $m$. 
In contrast, for a weakly superadiabatic stratification ($\delta > 0$), the frequencies of low-$m$ modes decrease (less prograde). 
This behavior originates from the phase relation between the entropy perturbation $s'$ and the radial velocity $v_r$. 
For $\delta < 0$, the phase of $s'$ is ahead of $v_r$ in longitude, so that buoyancy acts as an additional restoring force and enhances the prograde propagation. 
Conversely, for $\delta > 0$, $s'$ lags behind $v_r$, such that buoyancy partially offsets the restoring force associated with the compressional $\beta$-effect, thereby reducing the mode frequency.
Furthermore, \citet{bekki2022a} found that, when $\delta >0$, a critical azimuthal order $m_c$ exists above which the modes become linearly unstable ($\Im[\omega] > 0$).
These convectively unstable modes are characterized by a positive correlation between $v_r$ and $s'$, implying the radially upward heat transport. 
The value of $m_c$ decreases as $\delta$ increases. 
It was found that, once the modes become convectively unstable, their prograde frequencies are substantially reduced and become relatively insensitive to $\delta$. 
A similar critical-$m$ behavior has also been discussed in 1.5D cylindrical models \citep{hindman2023,jain2026}.
To date, no unambiguous detection of such prograde columnar modes has been established in the Sun. 
Recently, prograde signals potentially associated with these modes have been reported using helioseismic normal-mode coupling \citep{hanasoge2026,mandal2026}.
However, their physical nature and mode identification remain to be established before they can be used in forward modeling to constrain $\delta$.

As already noted, the HFR modes are strongly non-toroidal and therefore are also expected to be sensitive to the superadiabaticity $\delta$. 
\citet{bekki2024a} demonstrated that the observed frequencies of the HFR modes can be used as a diagnostic of $\delta$. 
In that study, it was first shown that the HFR-mode frequencies depend only weakly on the turbulent viscosity and on the near-surface superadiabaticity, but instead sensitively depend on the value of $\delta$ in the bulk of the convection zone ($r \leq 0.95R_{\odot}$). 
The sensitivity of the mode frequencies to $\delta$ can be understood in terms of the same physical mechanism discussed above: buoyancy either enhances or weakens the restoring force, depending on the sign and magnitude of $\delta$.
Under the observed differential rotation profile, \citet{bekki2024a} found that the measured frequencies of the HFR modes are best reproduced in the linear model when the bulk convection zone is slightly subadiabatic, with $-5.0\times10^{-7} \leq \delta \leq -2.5\times10^{-7}$.
Although this level of subadiabaticity is very small, the corresponding Brunt-Väisälä frequency near the base of the convection zone is comparable to the solar rotation rate ($N/\Omega \sim 0.5-0.75$).
This implies that buoyancy may provide a non-negligible restoring force for the HFR modes in addition to the Coriolis force, suggesting that they may be regarded as gravito-inertial modes \citep[][]{bekki2024a}.

\citet{gizon2021} and \citet{dey2024iau} showed that the $m=1$ high-latitude mode also exhibits a dependence on the superadiabaticity $\delta$. 
The properties of the $m=1$ high-latitude mode are primarily governed by baroclinic instability. 
Although the mode has almost negligible dependence on $\delta$ for $\delta < 0$, it was found that the spiral structure of the surface eigenfunction reverses its orientation relative to that observed on the Sun when $\delta > 2.0\times10^{-7}$.
This behavior likely reflects a competition between entropy perturbations driven by the latitudinal entropy gradient and those driven by the superadiabaticity (Dey et al. in prep.). 

Taken together, these results suggest that superadiabaticity in the deep solar convection zone is much closer to adiabatic, or even slightly subadiabatic, than traditionally estimated by mixing-length theory \citep[e.g.,][]{vitense1953,bohm1958}, challenging the standard paradigm of convective energy transport in the Sun.
In fact, this picture is broadly consistent with several recent numerical simulations of solar-like turbulent convection, in which the lower convection zone becomes weakly subadiabatic \citep{kapyla2017,bekki2017b,nelson2018,hotta2022,noraz2025}. 
This weakly subadiabatic layer, often referred to as the \textit{Deardorff} layer \citep{brandenburg2016}, is thought to be maintained by non-local heat transport by strong downflow plumes.

Finally, we note that most of these previous studies have assumed a spatially uniform $\delta$ for simplicity. 
Future work should therefore pursue forward modeling that incorporates a radially varying profile of $\delta$ (Dey et al., in preparation).

\begin{figure}[t]
\centering
\includegraphics[width=0.7\textwidth]{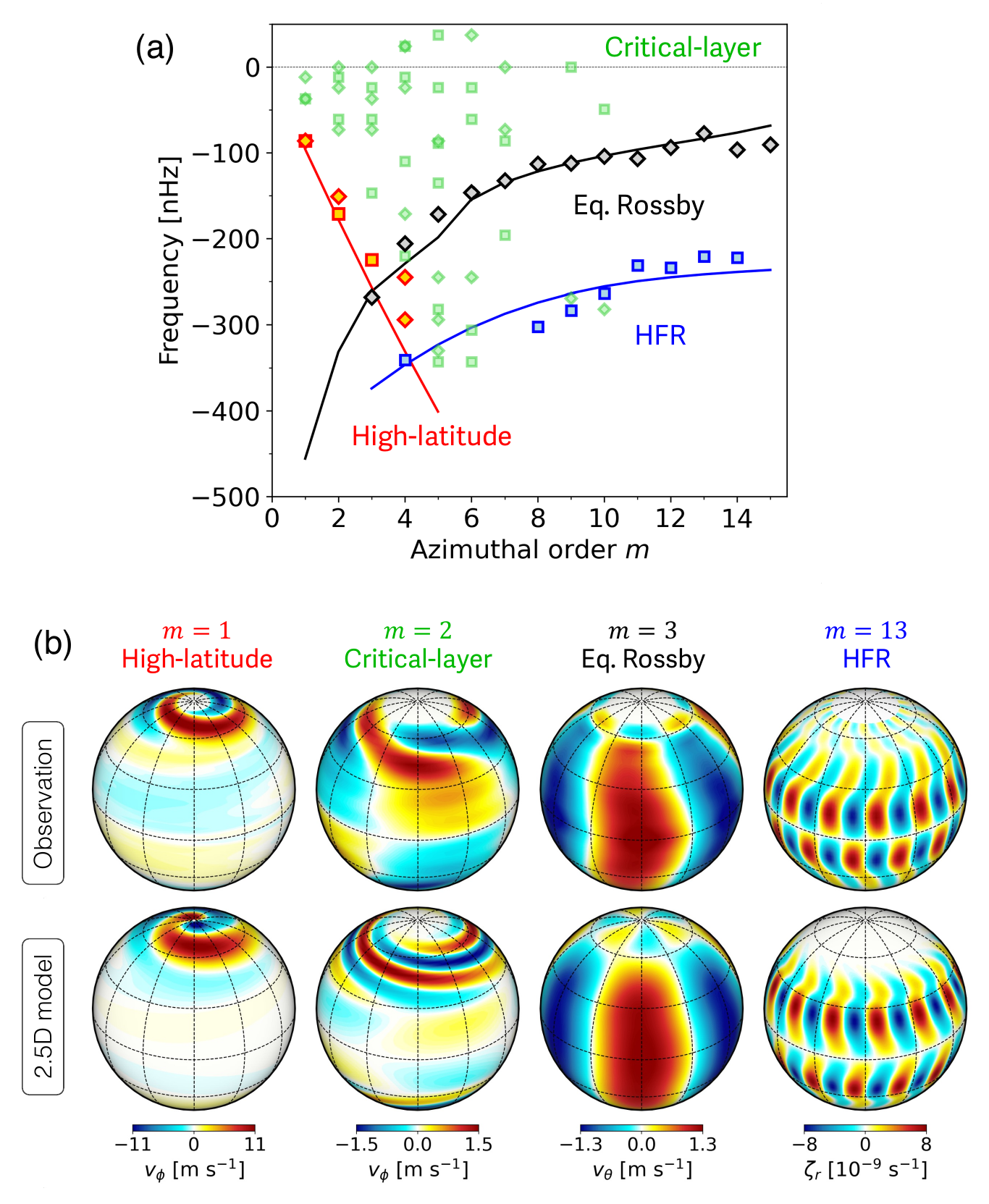}
\caption{
Observations and linear modeling of the solar inertial modes.
(a) Frequencies of the observed modes measured in the Carrington frame ($\Omega_{\rm Car}/2\pi=456$~nHz).
The symbols show the observed mode frequencies \citep[from][]{loeptien2018,gizon2021,hanson2022}, where the modes with north-south symmetric and antisymmetric radial vorticity are denoted by diamonds and squares.
The curves show the theoretical dispersion relations computed using the 2.5D linear model \citep[from][]{bekki2024a}, after fixing the coding error reported by \citet{dey2026a}.
The linear model includes the observed differential rotation, latitudinal entropy gradient, weakly subadiabatic convection zone ($\delta = -5 \times 10^{-7}$), and turbulent viscosity which varies as $\propto \rho_0^{-1/2}$ normalized such that the volume-averaged value is $5 \times 10^{12}$~cm$^2$~s$^{-1}$.
Black, red, blue, and green colors represent the equatorial Rossby modes, high-latitude modes, high-frequency retrograde (HFR) modes, and critical-layer modes.
(b) Surface eigenfunctions of the selected inertial modes, comparing the observations (top panels) and the linear model (bottom panels).
}\label{fig:gizon21_comp}
\end{figure}

\subsection{Comparison of Linear Inertial Modes with Solar Observations} \label{sec:summary_linear}

Figure~\ref{fig:gizon21_comp} compares the frequencies and eigenfunctions of the modes observed on the Sun \citep[][]{loeptien2018,gizon2021,hanson2022} with those of linear eigenmodes computed using the 2.5D model with solar differential rotation and latitudinal entropy gradient \citep[][]{bekki2022a,bekki2024a}.
The linear model incorporating all the realistic effects (differential rotation, latitudinal entropy gradient, weakly subadiabatic convection zone, and moderate turbulent diffusivities) is found to reproduce reasonably well the observed frequencies of the equatorial Rossby modes, the high-latitude modes, and the HFR modes. 
Except for the high-latitude modes, all modes shown here are linearly stable. 
For the equatorial Rossby modes, we show the $n=0$ modes rather than the $n=1$ mixed modes. 
At high $m$, these modes tend to localize near the base of the convection zone due to the presence of viscous critical layers. 
The baroclinically unstable high-latitude modes successfully reproduce both the nearly non-dispersive frequencies and the characteristic spiral structure observed on the Sun. 
The critical-layer modes reported by \citet{gizon2021} are distributed over a range of $m$ and frequencies and cannot be described by a single well-defined dispersion relation. 
As illustrated by the $m=2$ case in Fig.~\ref{fig:gizon21_comp}b, the linear calculations capture the oscillatory behavior near the critical layers, although noticeable differences remain in the detailed flow structures. 
Finally, the surface eigenfunctions of the HFR modes are roughly consistent with observations, with minor differences in their detailed structure.
As already noted, linear eigenmodes are sensitive to various model parameters. 
In particular, \citet{gizon2021} pointed out that the eigenfunctions of high-latitude and critical-layer modes do not match the observed ones when the superadiabaticity $\delta$ is strongly positive or when the turbulent viscosity $\nu_{\rm t}$ is too large.
Quantifying these parameter dependencies and resolving such discrepancies will be important future work.

\section{Nonlinear Models and Mode Amplitudes} \label{sec:nonlinear} 

Although linear models have played a central role in identifying and classifying solar inertial modes detected in observations, linear theory alone is insufficient to address key questions regarding the excitation, saturation, and amplitudes of inertial modes, as well as their potential dynamical impact on the solar interior. 
Addressing these issues requires nonlinear models.
Once the nonlinear terms in the governing equations are retained, however, a fundamental complication arises:
In rotating systems, the nonlinear interactions of flows inevitably lead to the transport of thermal energy and angular momentum in preferred directions. 
As a consequence, the background profiles of entropy or differential rotation -- commonly prescribed a priori as crucial ingredients in linear models --
can no longer be maintained.

Within the nonlinear regime, the most straightforward approach is to analyze inertial modes in fully nonlinear simulations of rotating convection, in which their excitation, saturation, and amplitudes are determined self-consistently \citep[][]{bekki2022b,blume2024,fuentes2026}. 
Such simulations are commonly referred to as \textit{first-principles} models, as the mean flows are generated as a consequence of the resolved dynamics.
While these first-principles simulations have provided important insights into the nonlinear nature of solar inertial modes, they face intrinsic limitations in maintaining the background states that are consistent with solar observations.

\subsection{Inertial Modes in Rotating Convection Simulations}

\begin{figure}[t]
\centering
\includegraphics[width=0.975\textwidth]{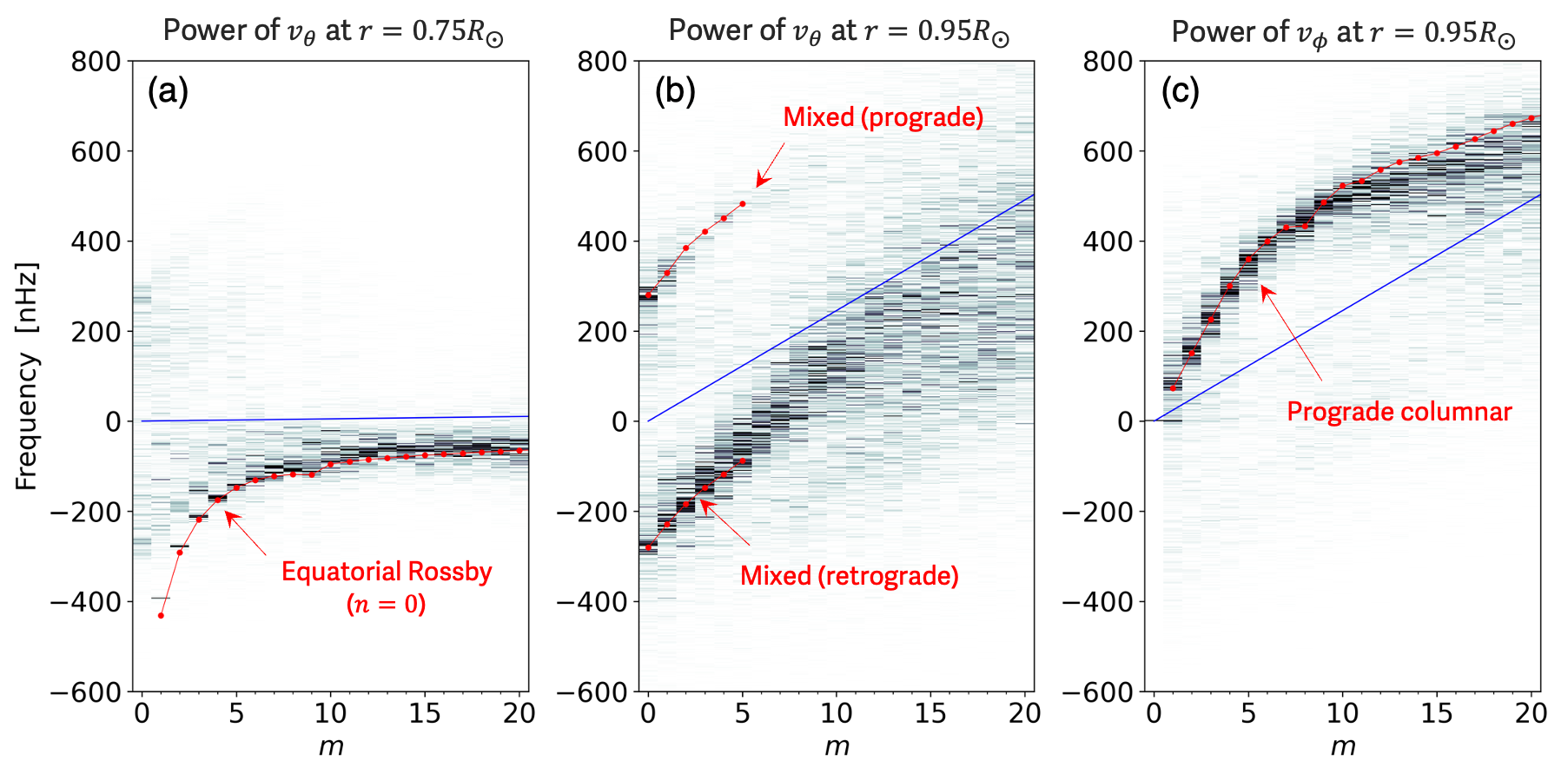}
\caption{
Equatorial power spectra of horizontal velocity from fully nonlinear rotating convection simulation of \citet[][]{bekki2022b}, in which the generated differential rotation is weaker and closer to the Taylor-Proudman state than in the Sun.
(a) Power spectrum of latitudinal velocity $v_{\theta}$ near the base of the convection zone ($r=0.75R_{\odot}$), averaged over the equatorial band ($\pm 15^{\circ}$), showing the ridge corresponding to the equatorial Rossby modes.
(b) Same as panel a but near the surface ($r=0.95R_{\odot}$), showing the mixed inertial modes between $\lambda=1$ prograde columnar modes and $n=1$ equatorial Rossby-like modes.
(c) Same as panel b but for the longitudinal velocity $v_{\phi}$, showing the ridge corresponding to $\lambda=0$ prograde columnar modes.
In all panels, the power is normalized at each $m$.
The blue curves show the equatorial rotation rate at each height, $\omega = m[\Omega(r,\pi/2)-\Omega_0]$.
Overplotted in red represents the dispersion relations computed from the linear model that takes into account the differential rotation profile from the nonlinear simulation.
}\label{fig:nonlin}
\end{figure}

\citet{bekki2022b} carried out the first systematic analysis of inertial modes in fully nonlinear simulations of solar-like rotating convection in a spherical shell.
In their simulations, a reduced luminosity was used to circumvent the so-called convective conundrum  \citep[e.g.,][]{hotta2023review,kapyla2023review} and to establish the solar-like differential rotation (with faster equator and slower poles).
In these simulations, inertial modes are primarily stochastically excited and damped by turbulent convection, while instabilities associated with shear and baroclinicity play only a minor role, owing to the relatively weak differential rotation and the very small baroclinicity established in the model.
Long time series of the simulations were analyzed, and the coherent large-scale modes at low frequencies embedded in turbulent convection are extracted using singular value decomposition \citep[SVD, see also][]{proxauf2020} and the mode amplitudes are measured.
\citet{bekki2022b} has identified the prograde columnar modes with no latitudinal node ($\lambda=0$) in axial vorticity $\zeta_z$ as the most dominant modes, transporting substantial amount of heat ($\sim 60\%$ of what is required) and angular momentum ($\sim 40\%$ of the net Reynolds stress) near the equator (Fig.~\ref{fig:nonlin}c).
The equatorial Rossby modes with no radial nodes ($n=0$) have also been found and successfully identified in the power spectra near the base of the convection zone (Fig.~\ref{fig:nonlin}a).
\citet{bekki2022b} reported that they have the near-surface velocity amplitudes of a few m~s$^{-1}$ and the linewidths of about $20-30$~nHz, which are roughly comparable to the observed values.
This implies that such modes can be stochastically excited to realistic amplitudes by turbulent convection.
Furthermore, they confirmed the existence of the \textit{mixed} inertial modes between the retrograde equatorial Rossby-like modes with one radial node ($n=1$) and the prograde columnar modes with one latitudinal node ($\lambda=1$) in $\zeta_z$ (Fig.~\ref{fig:nonlin}b).
For all of these modes, \citet{bekki2022b} compared the mode extracted from the nonlinear simulations with those obtained from the 2.5D linear eigenmode calculations which explicitly takes into account the differential rotation profile from the nonlinear simulation.
It was shown that the spatial eigenfunctions and the frequencies of the inertial modes present in the fully nonlinear simulations were in good agreement with those predicted by the corresponding linear models.
This result indicates that, at least within the parameter regime of \citet{bekki2022b}, the inertial modes are largely described by linear theory, while the nonlinearity primarily affects their excitation, saturation, and amplitudes. 
This finding also provides a supporting evidence for the use of linear models in the identification of observed modes \citep[as has been done in][]{gizon2021} and inferences of the Sun's deep interior, as discussed in \S~\ref{sec:linear}.

Beyond the work of \citet{bekki2022b}, only a limited number of nonlinear simulations explicitly addressing solar inertial modes have been conducted. 
\citet{blume2024} reported the existence of various inertial modes in a nonlinear magneto-convection simulation including both the convection zone and the radiative interior.
Using the rotation rate three times faster than in the real Sun, the simulation has successfully achieved a solar-like differential rotation in the convection zone and the thin tachocline confined due to the local dynamo action \citep[][]{matilsky2022}.
In addition to the inertial modes identified by \citet{bekki2022b}, \citet{blume2024} also confirmed the presence of the HFR modes in their simulation.
\citet{blume2024} have further demonstrated that, in the uniformly rotating radiative interior, classical Rossby modes exist over a wide range of $(\ell,m)$ combinations, not limited to the sectoral modes.
These Rossby modes were shown to have frequencies different from those of the equatorial Rossby modes in the convection zone, and thus might live in a cavity distinct from that in the convection zone.

We note that both of the above studies \citep[][]{bekki2022b,blume2024} are based on simulations that did not reproduce the Sun in a quantitatively realistic manner.
Also, both studies focused on detailed analyses of individual simulations, without systematically exploring parameter dependencies.
These reflect the intrinsic difficulties associated with fully nonlinear simulations. 
In particular, the analysis of low-frequency inertial modes requires very long time series spanning a decade or more, which renders such studies computationally demanding.
Furthermore, the differential rotation is highly sensitive to changes in control parameters, making systematic parameter variations challenging.
In the future, however, it will be necessary to analyze nonlinear simulations carried out with more realistic solar parameters \citep[e.g.,][]{hotta2022,popovas2022,noraz2025,hotta2025}
as well as to conduct systematic parameter surveys to investigate how the excitation of inertial modes depend on dimensionless parameters such as the Rossby, Rayleigh, and Prandtl numbers \citep[e.g.,][]{fuentes2026}.

\subsection{Simulations of Baroclinically Unstable High-Latitude Modes}

\begin{figure}[t]
\centering
\includegraphics[width=0.95\textwidth]{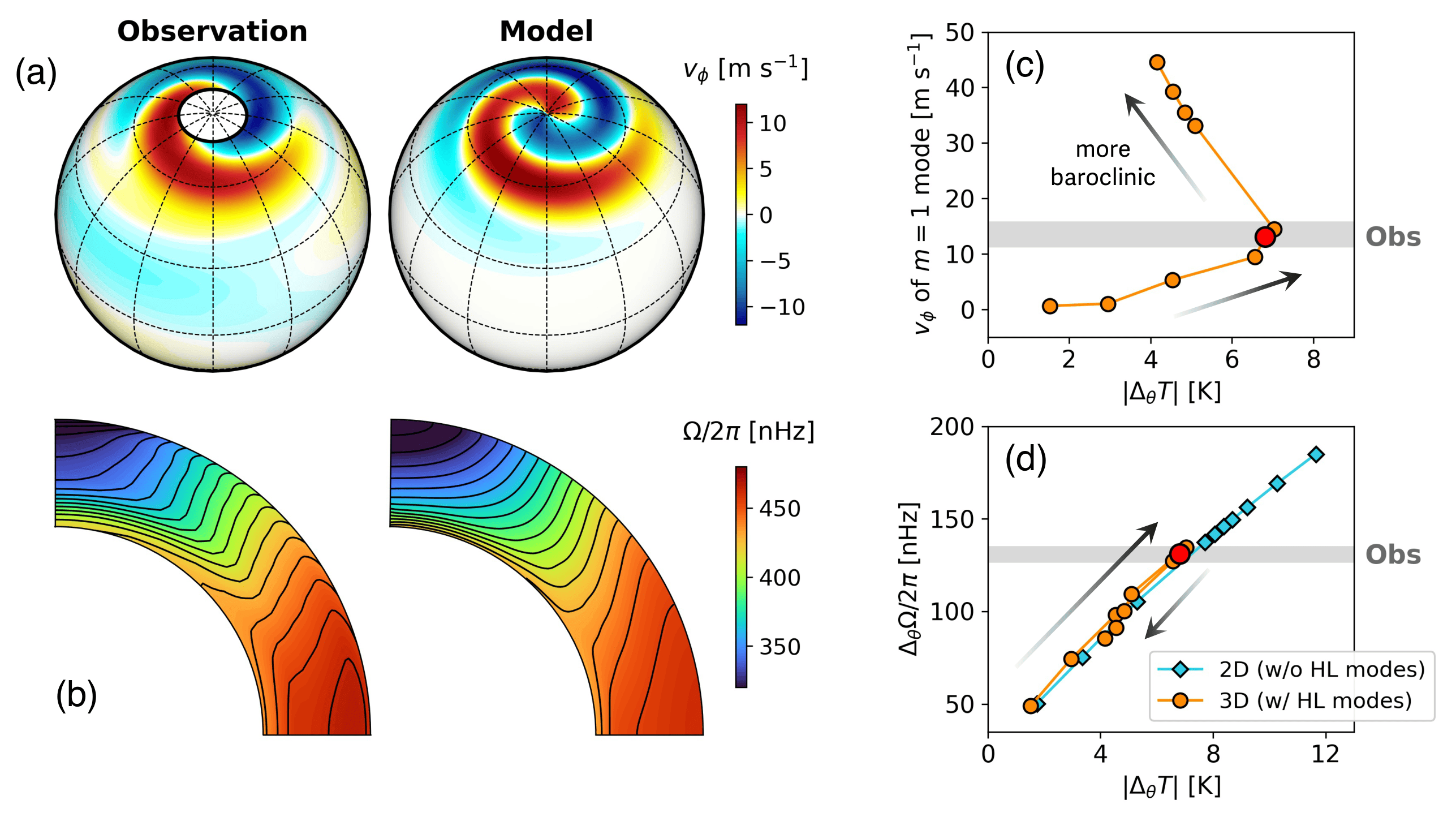}
\caption{
Results from 3D nonlinear mean-field simulations of the high-latitude inertial modes by \citet{bekki2024b}.
(a) Comparison of the near-surface longitudinal velocity eigenfunction of the $m=1$ high-latitude mode from the observation and from the mean-field model.
(b) Comparison of the differential rotation profile from the global helioseismology and from the model.
(c) Amplitude of the $m=1$ high-latitude mode as a function of the pole-to-equator temperature difference $|\Delta_\theta T|$ as the baroclinic forcing is increased (denoted by arrows). The gray shaded area shows the observed mode amplitude.
(d) Latitudinal differential rotation $\Delta_\theta \Omega$ as a function of $|\Delta_\theta T|$. For comparison, the results from the 2D axisymmetric model (where the non-axisymmetric modes are not present) are shown in cyan.
In panels a and b, the model corresponds to the best-fit case \citep[case 5 in][]{bekki2024b} denoted by red circles in panels c and d.
}\label{fig:meanfield}
\end{figure}

While various inertial modes at low to middle latitudes can be studied with first-principles models of rotating convection, the high-latitude modes are not nicely reproduced in these simulations \citep[see Appendix~C in][]{bekki2022b}.
In particular, their characteristic spiral patterns and large amplitudes of the observed high-latitude modes have not been reproduced. 
This discrepancy is likely attributable to the weak differential rotation and insufficient baroclinicity established in these simulations, both of which are expected to play a crucial role for exciting high-latitude modes \citep[][]{bekki2022a}.
In first-principles models, however, the differential rotation and the baroclinicity emerge self-consistently due to the nonlinear dynamics and are therefore difficult to control systematically.

To overcome this limitation, \citet{bekki2024b} adopted an alternative nonlinear approach based on a 3D mean-field framework (Fig.~\ref{fig:meanfield}). 
The mean-field framework has been widely employed in numerical studies of solar large-scale flows and dynamo processes. Traditionally, most implementations have been restricted to 2D axisymmetric models \citep[e.g.,][]{kitchatinov1995,dikpati1999,rempel2005}, but fully 3D formulations have increasingly been developed in recent years \citep[e.g.,][]{yeates2013,miesch2014,pipin2022,bekki2023}.
In this framework, small-scale convective motions are not explicitly resolved but their effects are instead parameterized through a prescribed $\Lambda$-effect \citep[e.g.,][]{rudiger1989,rempel2005}, by which the solar-like differential rotation is generated (see Fig.~\ref{fig:meanfield}b). 
This formulation allows the background baroclinicity to be controlled systematically as an input parameter, providing a controlled environment to investigate the amplitudes and the nonlinear saturation of baroclinically unstable high-latitude modes.
\citet{bekki2024b} conducted a series of 3D mean-field simulations designed to systematically vary the background baroclinicity in the convection zone.
It was demonstrated that, as the background baroclinicity is increased, the $m = 1$ high-latitude mode is baroclinically excited and its mode amplitude increases. 
As shown in Fig.~\ref{fig:meanfield}a, their mean-field simulations have reproduced key observational features of the high-latitude modes, including their spiral spatial structures and surface velocity amplitudes of order $10$~m~s$^{-1}$, in good agreement with the solar observations. 
The parameter study of \citet{bekki2024b} further revealed that the amplitude of the $m=1$ high-latitude mode is nonlinearly coupled with the pole-to-equator temperature difference $\Delta_\theta T = T_{\rm eq}-T_{\rm pole}$ in the convection zone.
As the baroclinicity is increased, both $|\Delta_\theta T|$ and the amplitude of the $m=1$ high-latitude mode increase accordingly.
However, once the mode amplitude exceeds the threshold value of $\sim 15$~m~s$^{-1}$, further increase in baroclinic forcing lead to a reduction in $|\Delta_\theta T|$ (see Fig.~\ref{fig:meanfield}c).
This behavior represents a nonlinear negative feedback mechanism and can be attributed to the equatorward heat transport by the high-latitude modes, which prevents $|\Delta_\theta T|$ from growing without bound.
In turn, the reduced $|\Delta_\theta T|$ suppresses further baroclinic excitation of the high-latitude modes, leading to their saturation.
From their parameter survey, \citet{bekki2024b} inferred that reproducing the observed amplitudes requires $|\Delta_\theta T| \sim 7$~K in the middle convection zone.
This estimate is consistent with the value predicted by thermal wind balance \citep[][]{thompson2003,miesch2006}.
We also note that this $|\Delta_\theta T| \sim 7$~K is extremely small compared to the background temperature in the Sun's convection zone, and is therefore very hard to detect by direct observations \citep[e.g.,][]{rast2008}.
The newly-identified high-latitude inertial modes provide a means to probe and indirectly constrain $|\Delta_\theta T|$ from surface observations.

The study of \citet{bekki2024b} also has important implications for the Sun's differential rotation.
Figure~\ref{fig:meanfield}d compares the latitudinal differential rotation $\Delta_\theta \Omega = \Omega_{\rm eq}-\Omega_{\rm pole}$ from the 2D axisymmetric model and from the 3D (non-axisymmetric) model.
In 2D model where the effects of baroclinic modes are excluded, $\Delta_\theta \Omega$ can be increased arbitrarily by tuning model parameters.
In contrast, in a realistic 3D model, $\Delta_\theta \Omega$ cannot exceed a theoretical upper bound set by the nonlinear feedback of the high-latitude modes, which regulates $|\Delta_\theta T|$ to be less than $\sim 7$~K. 
This is due to the poleward angular momentum transport by the meridional circulation induced by the reduction in $|\Delta_\theta T|$.
Remarkably, the Sun's differential rotation appears to operate close to this theoretical upper bound, implying that high-latitude modes may play a significant dynamical role in regulating the Sun’s latitudinal differential rotation.

We note that the numerical model of \citet{bekki2024b} was designed to isolate the role of the high-latitude modes and their interplay with the background baroclinicity within a controlled framework.
This simplification comes at the cost of ignoring additional physical processes such as the effects of magnetic fields and turbulent convection, which may also influence the excitation and nonlinear saturation of the high-latitude modes.
Incorporating these realistic effects in future models will be essential for assessing the robustness of their results and conclusions.

We also note that \citet{souzagomes2025} presented a series of nonlinear simulations of non-convecting fluid in a global spherical shell in which the differential rotation is forced through a body-force term that relaxes the azimuthal velocity toward a prescribed target profile.
By controlling the shear amplitudes of the established differential rotation, they found that the high-latitude modes can be self-excited as the imposed shear was increased.

\subsection{Other Approaches to Estimate Mode Amplitudes}

The method introduced by \citet{philidet2023} provides a simple and tractable framework to model the excitation and damping of linearly stable inertial modes by turbulent convection. In this framework, the linearized wave equations  are driven by stochastic sources and incorporate dissipation through turbulent viscosity. Rather than computing individual realizations, the approach predicts the expectation value of the oscillation power spectrum. This quantity depends on two key ingredients: the Green’s function of the system and the covariance of the turbulent sources. A parametric model of the source covariance can be adjusted to constrain the excitation mechanism. The Green’s functions can be computed either through a normal-mode expansion \citep{philidet2023} or by direct numerical solution of the governing equations \citep[see, e.g.,][]{Pham2025, Faucher2025}. Conceptually, this approach follows the stochastic excitation theory developed for solar acoustic modes by \citet{Goldreich1990}, which itself builds on \citet{Lighthill1952}’s theory of sound generation by turbulence. In this framework, linearly unstable modes must be ``discounted" by considering the system dynamics over a finite time horizon \citep{Jovanovic2004}. In their 2D study, \citet{philidet2023} demonstrated that linearly stable Rossby modes can be excited by the vorticity of turbulence to reach surface velocity amplitudes of order $1$~m~s$^{-1}$, comparable to observed values. This result supports the hypothesis that linearly stable inertial modes are excited by turbulent convection, in agreement with conclusions drawn from numerical simulations. 
Another important result from \citet{philidet2023} is that peaks of excess power in the spectrum may arise from the superposition of contributions from multiple modes. In this regime, the power spectrum is dense, meaning that the frequency spacing between modes can be smaller than their linewidths. As a result, individual modes are not fully resolved and the observed power can be described as a sum of overlapping Lorentzian profiles.

Nonlinear excitation of Rossby modes may also occur through an inverse energy cascade. 
It has long been known that the transition from turbulence-dominated to Rossby-wave-dominated dynamics occurs near the Rhines scale \citep[][]{rhines1975}.
In the solar context, \citet[][]{loeptien2018} discussed the observed spatial scales of the equatorial Rossby modes in terms of the Rhines scale.
More recently, using shallow-water simulations, \citet{dikpati2022} showed that kinetic energy injected at small spatial scales can be transferred to global Rossby modes. Although this model is restricted to a shallow-water framework, it provides a complementary mechanism for the nonlinear excitation of the large-scale linearly stable modes.

For linearly unstable modes, however, the above linear framework is not applicable, since their amplitudes are determined by nonlinear saturation. In this case, weakly nonlinear theory may be used to relate the equilibrium amplitudes of unstable modes to their linear growth rates.
Using fully nonlinear simulations of turbulent motions on a two-dimensional differentially rotating sphere, \citet{mushtaq2026} showed that, in the vicinity of marginal instability, the nonlinear saturation of the $m=1$ high-latitude mode follows a supercritical Hopf bifurcation that can be captured by weakly nonlinear theory \citep[using the expansion of][]{Stuart1960}.
They also pointed out that the mode shape in the nonlinear saturated state  includes   higher-order harmonics $m \ge 2$, which may have important implications for the interpretation of the observed power spectra.
Although the study by \citet{mushtaq2026} provides valuable physical insight into the nonlinear saturation of inertial mode, it should be noted that the unstable modes in this model are excited by shear instability and saturate via their Reynolds stress feedback. 
Extending the model to a more realistic 3D configuration, where baroclinic instability plays a more prominent role, will be important.

Compared with first-principles models of rotating convection, these simplified nonlinear approaches sacrifice dynamical self-consistency in exchange for analytical clarity and computational efficiency. 
They provide valuable insight into specific aspects of inertial-mode excitation and saturation, while complementing more comprehensive nonlinear models.

\section{Discussion and Future Prospects} \label{discussion}

Observational and theoretical studies of solar inertial modes have advanced rapidly in recent years. 
These developments are expected to provide a new framework for probing the solar interior in ways that complement traditional p-mode helioseismology. 
Nevertheless, a significant gap still remains between current observations and theoretical models. 
In this section, we briefly summarize the outstanding issues and outline prospects for future modeling efforts.

\subsection{Remaining Gaps and Open Issues}\label{sec:open}

\subsubsection*{-- Prograde Columnar Modes}

The most significant discrepancy between theoretical models and observations concerns the existence of prograde columnar (or thermal Rossby) modes. 
Both linear calculations and nonlinear simulations consistently predict the presence of dominant prograde columnar modes. 
However, their unambiguous identification in solar observations remains challenging.
Recently, \citet{hanasoge2026} and \citet{mandal2026} reported candidate detections of such modes using a normal-mode coupling technique to infer subsurface flow spectra.
Although they claimed evidence for prograde power ridges in the Carrington co-rotating frame, their measured frequencies are substantially lower than those predicted for prograde columnar modes in the linear models at low azimuthal orders.
A quantitative comparison with theoretical models is therefore needed to establish the physical nature of the detected signals and, in particular, whether they correspond to convectively driven modes.

The existence of the large-scale prograde columnar modes is closely related to the so-called solar convective conundrum \citep[e.g.,][]{hotta2023review}.
Once observed, their mode properties could provide diagnostics of superadiabaticity $\delta$ and the convective Rossby number $\mathrm{Ro}$ in the Sun's deep interior.
\citet{hanasoge2026} and \citet{mandal2026} argued that their inferred small mode amplitudes ($\sim 10~\mathrm{cm\,s^{-1}}$ for toroidal and $\sim 4~\mathrm{m\,s^{-1}}$ for poloidal flow motions at $r<0.98R_{\odot}$) imply that the deep solar interior is strongly rotationally constrained (i.e., at very small Ro).
However, caution is needed when relating the inferred mode amplitudes directly to the convective Rossby number, since the mode velocity is not a direct measure of the characteristic convective velocity.
Indeed, coherent columnar patterns tend to break down at large Ro \citep[e.g.,][]{featherstone2016}, so that weak mode amplitudes do not necessarily imply a small Ro.
Addressing this issue requires systematic investigations of the amplitudes of prograde columnar modes in fully nonlinear rotating convection simulations.

\subsubsection*{-- Equatorial Rossby Modes}

Although equatorial Rossby modes are the best-observed class of solar inertial modes, they are not yet fully understood from a theoretical perspective. 
In particular, linear models yield both $n=0$ and $n=1$ modes within nearly the same frequency range for $m \gtrsim 5$. 
Because these modes exhibit similar surface eigenfunctions, it remains difficult to determine whether the observed modes correspond to the $n=0$ branch, the $n=1$ branch, or a hybrid of both \citep[][]{bekki2022a}.
For $m=3$, the mode is widely believed to correspond to the $n=0$ branch \citep[][]{gizon2021}. 
However, this mode lies close in frequency to the $m=3$ high-latitude mode, so the observed high-latitude structure may contain contributions from the nearby branch \citep[][]{mandowara2024IAU}.
A careful assessment of the latitudinal eigenfunction structure is therefore required both from observational and theoretical sides.

Long-term observational studies have revealed significant solar-cycle variations in both the amplitudes and frequencies of equatorial Rossby modes \citep[][]{waidele2023,lekshmi2026}. 
Recent measurements further suggest that, at large azimuthal order ($m \gtrsim 20$), these modes exhibit a frequency splitting into lower- and higher-frequency branches \citep[][]{hanasoge2026b}.
Such behavior has been interpreted as indicative of magneto-Rossby modes, in which the Lorentz force acts as an additional restoring force \citep[][]{zaqarashvili2007}. 
These findings strongly suggest that realistic modeling of equatorial Rossby modes must incorporate magnetic effects in fully MHD models (\S~\ref{sec:mhd}).

\subsubsection*{-- High-Latitude Modes}

High-latitude modes, which exhibit the largest amplitudes among the observed inertial modes, have been the subject of active theoretical investigation in recent years. 
In particular, within a framework of baroclinic instability, both linear models and nonlinear mean-field simulations have been shown to reproduce remarkably well the observed spatial structure, frequencies, and amplitudes of the high-latitude modes \citep[][]{bekki2022a,bekki2024b}, thereby providing a plausible theoretical foundation for their origin. 
Despite these successes, key questions remain open.
One example concerns the amplitudes of  observed high-latitude modes: the $m=1$ mode with antisymmetric $v_\phi$ has a much larger amplitude than its symmetric counterpart \citep[][]{gizon2021,liang2025}, while linear models predict nearly identical growth rates for modes with both symmetries.
The physical mechanism responsible for maintaining this difference remains unclear.
Furthermore, in the nonlinear mean-field model of \citet{bekki2024b}, the effects of turbulent convection and superadiabatic stratification were ignored. 
Future work will need to assess the robustness of these previous results to the inclusion of such realistic effects.
An alternative explanation invokes MHD instabilities as the origin of high-latitude modes \citep[][]{dikpati2024}.

Recent observations further indicate that the amplitudes of high-latitude modes also exhibit significant solar-cycle dependence \citep[][]{gizon2024iau,liang2025,ding2025,lekshmi2026}. 
These observations underscore the need for MHD models for the high-latitude modes. 
In general, magnetic fields are known to suppress baroclinic instability \citep[][]{gilman2015,gilman2017}. 
Because baroclinically unstable high-latitude modes can modulate the differential rotation, they may  interact with the high-latitude branch of the torsional oscillation \citep[e.g.,][]{howe2013}.
Furthermore, the $m=1$ mode has a characteristic pole-crossing flows, as confirmed both observationally \citep[][]{ulrich2001,liang2025} and theoretically \citep[][]{bekki2024b}. 
How such pole-crossing flow impacts the surface flux transport and the internal dynamo process remains an open and worthwhile question for future investigation \citep[e.g.,][]{heinemann2025,chitta2025}.

\subsection{Directions for Improved Modeling}

Solar inertial modes differ fundamentally from the p modes that have conventionally been used in helioseismology in several key respects. 
First, they are intrinsically 3D phenomena. 
With solar differential rotation, most inertial modes cannot be represented by spherical harmonics and instead exhibit genuinely 3D eigenfunctions. 
Second, some large-amplitude modes can feed back onto the mean state, implying that nonlinear effects must be taken into account for these modes \citep[][]{bekki2024b}. 
Third, inertial modes are sensitive to magnetic fields and exhibit significant solar-cycle variations \citep[e.g.,][]{lekshmi2026}. 
A comprehensive understanding of such modes, therefore, requires the development of sophisticated numerical models.
There are several promising directions in this regard, as outlined below.

\subsubsection{Code Validation and Benchmarking}

Before outlining future directions, we would like to emphasize the importance of cross-checks between different numerical codes. 
Several 2.5D eigenvalue solvers for solar inertial modes have been developed using different numerical schemes and boundary conditions, each with their own strengths and limitations. 
Systematic benchmark calculations between such codes are essential both to ensure numerical reliability and to understand quantitative differences arising from implementation details. 
In particular, a detailed comparison between the code presented in \citet{bekki2022a} and the open-source framework \texttt{Dedalus} \citep{mukhopadhyay2025} revealed a coding error in the implementation of the turbulent diffusion terms in the former, as reported by \citet{dey2026a}.
The coding error effectively enhanced the turbulent viscosity near the surface, thereby affecting the mode damping rates and, at sufficiently large viscosity, also the mode radial structures. 
However, these corrections do not alter the main physical conclusions of \citet{bekki2022a}.
We also note that all linear calculations presented in this review paper are based on a corrected version of the code developed by \citet{bekki2022a}.

\subsubsection{Including the Radiative Interior and the Near Surface Layer} \label{sec:domain}

So far, we have primarily focused on inertial modes trapped within the convection zone. 
In such models, the numerical domain is typically restricted to $r \in [0.71R_{\odot},\,0.985R_{\odot}]$, with impenetrable and stress-free boundary conditions imposed at both radial boundaries. 
Recently, \citet{mukhopadhyay2026} extended the lower boundary down to $0.5R_{\odot}$ to investigate the effects of the subadiabatic radiative zone on inertial modes. 
They showed that, although the presence of the radiative zone modifies the eigenfunctions near the base of the convection zone and the mode damping rates through penetration into the overshoot region, its impact on the mode frequencies and the surface eigenfunctions is largely negligible. 
Furthermore, they demonstrated that, in the subadiabatic radiative interior, buoyancy maintains the radial force balance, allowing classical Rossby modes with any combination of $(\ell,m)$ to exist. 
Some of these modes were found to exhibit mixed behavior with convection-zone inertial modes.
The excitation mechanisms and amplitudes of such mixed Rossby-inertial modes remain to be explored in future nonlinear simulations with the radiative interior \citep[][]{blume2024}. 
It should also be noted that, for numerical feasibility, the overshoot layer adopted by \citet{mukhopadhyay2026} was significantly thicker than predicted by the Model S. 
Quantifying the influence of the radiative zone in models with a more realistic, stiffer overshoot layer therefore remains an important task for future studies.
If they have significant effects, inertial mode observations can be used to provide additional constraints on the thickness of the tachocline and overshoot region.

In the near-surface layer (above $0.985R_{\odot}$), the superadiabaticity $\delta$ and background stratification vary sharply, with density and pressure dropping by several orders of magnitude.
The numerical difficulty of resolving this layer has led many studies to exclude it from their numerical domains.
This treatment appears to have a limited impact on most inertial modes because the mode kinetic energy is typically concentrated in the lower convection zone and remains small near the surface \citep[see Fig.~2 in][]{gizon2021}.
Moreover, models employing free-surface boundary conditions yield results that closely resemble those obtained with impenetrable boundaries \citep[][]{mukhopadhyay2026}.
Nevertheless, because high-precision observations are largely confined to the near-surface layers, quantitative comparisons of velocity amplitudes require models that explicitly include this near-surface layer. 
Furthermore, this layer is essential for understanding mode excitation by turbulent convection or by the active region flows.
Future efforts should therefore extend current models to include the realistic surface layer using advanced numerical techniques \citep[e.g.,][]{saad2011}.

\subsubsection{Extension to MHD Models} \label{sec:mhd}

Several previous studies have considered magnetic effects indirectly. 
For example, \citet{goddard2020} investigated how sectoral Rossby modes respond to torsional oscillations associated with the solar cycle. 
The resulting cycle dependence appears to be inconsistent with observations \citep[][]{waidele2023,liang2025,lekshmi2026}, suggesting that magnetic fields have a more direct impact on inertial modes. 
Therefore, MHD models that explicitly incorporate magnetic fields are needed.

A straightforward extension is to insert an internal toroidal magnetic field as a background in the 2.5D linear model. 
Such a framework would enable us to address a number of key questions: 
How are the dispersion relations of inertial modes modified as a function of the strength and spatial distribution of the internal toroidal field? 
Do Lorentz forces induce additional frequency branches or splittings in the inertial-mode spectrum, similar to those predicted for magneto-Rossby modes \citep[e.g.,][]{zaqarashvili2007}?
Under what conditions on the internal magnetic field are shear and baroclinic instabilities suppressed \citep[e.g.,][]{gilman2015}?
Under what conditions do qualitatively new instabilities, such as magneto-shear instabilities \citep[e.g.,][]{gilman1997,dikpati2003} or the magneto-rotational instability \citep[e.g.,][]{kagan2014,vasil2024}, become dominant? 
These issues are directly relevant for interpreting recent observational findings.

MHD extensions are also essential in the nonlinear regime, which enables us to investigate the role of inertial modes in the dynamo process itself. 
Representative questions include the following. 
To what extent do the pressure perturbations associated with inertial modes affect the stability, storage, and subsequent buoyant rise of toroidal flux tubes \citep[e.g.,][]{fan2021}? 
How much can inertial modes with substantial kinetic helicity \citep[Fig.~S15 in][]{bekki2024b} contribute to the turbulent $\alpha$-effect? 
Can nonlinear interactions among magnetic fields, baroclinic instability, and large-scale mean flows account for the observed cycle dependence of torsional oscillations and high-latitude modes?

Finally, it is worth noting that magnetic effects are not limited to those associated with the large-scale dynamo. 
Recent numerical simulations suggest that a small-scale dynamo may operate vigorously in the solar convection zone \citep[e.g.,][]{hotta2022,warnecke2023}. 
In this context, \citet{bekki2025} numerically demonstrated that the small-scale Maxwell stresses can modify the dominant force balance and thereby alter the properties of prograde columnar modes in the rapidly rotating regime. 
Whether a similar mechanism operates in more slowly rotating regimes or for retrograde inertial modes remains an open question. 
Addressing these possibilities will be important for assessing the robustness of purely hydrodynamic interpretations of solar inertial modes.

\subsubsection{Towards Inertial-Mode Helioseismology} \label{sec:i-helio}

In p-mode global helioseismology, linear inversions based on observed mode frequencies have successfully constrained the internal sound-speed and rotation profiles \citep[e.g.,][]{christensen2002,thompson2003}. 
Such linear inversion techniques are  potentially applicable to inertial modes. 
For inertial modes, not only the frequencies but also the spatial eigenfunctions are model-dependent observables. 
However, establishing a one-to-one correspondence between computed linear eigenmodes and observations is often non-trivial, particularly when modes overlap in spectral space and avoided crossings occur.
Therefore, forward modeling is the most feasible approach at present.
To date, most forward-modeling efforts have focused on the modes that are best characterized by observations. In the future, a unified framework for quantitative fitting of the full power spectrum will be required, incorporating a model for the stochastic excitation of the modes \citep{philidet2023}.
A central difficulty is the large dimensionality of the model parameter space and the inherently nonlinear dependence of observables on those parameters. 
A key to progress is reducing the dimensionality of the model space through physically motivated parameterizations, together with inference frameworks that extend beyond purely linear inversions.

\section{Concluding Remarks} \label{conclusion}

Inertial modes have emerged as a new and powerful diagnostic of the solar convection zone, made accessible by long-duration observations  from SDO/HMI, GONG, and MWO \citep[notably][]{loeptien2018,gizon2021,waidele2023, liang2025, lekshmi2026}.
Their theoretical understanding has relied heavily on numerical modeling, because analytical approaches are severely limited, particularly when realistic differential rotation is included (\S~\ref{sec:diffrot}).
Over the past few years, linear eigenmode analyses in spherical-shell geometries have provided a coherent theoretical framework that reproduces the frequencies and eigenfunctions of the observed classes of inertial modes --- equatorial Rossby, high-frequency retrograde, critical-layer, and high-latitude modes --- and clarifies the roles of differential rotation, turbulent viscosity, baroclinicity, and superadiabaticity in shaping their properties
 \citep[][]{gizon2021,bekki2022a,bekki2024a}. 
Nonlinear models further indicate that the $m=1$ high-latitude inertial mode transports heat and angular momentum and may play a dynamical role in regulating the Sun’s differential rotation \citep[][]{bekki2024b}. 
Despite this progress, important gaps remain, such as definitive explanations for the mode amplitudes and linewidths (including the small amplitudes of the elusive prograde columnar modes), as well as the incomplete treatment of magnetic effects and realistic convective turbulence.
Bridging these gaps will require further advances in observations, eigenvalue solvers, and nonlinear simulations.


\begin{acks} 
We gratefully acknowledge frequent scientific discussions with many members of the Solar and Stellar Interiors Department at the MPS, including  R.H.~Cameron, P.~Dey, D.~Fournier, N.~Joshi, Z.-C.~Liang,  Y.~Mandowara, S.~Mukhopadhyay, and M.~Mushtaq. 
We are especially grateful to P.~Dey and S.~Mukhopadhyay for carrying out benchmark calculations of the linear models.
We additionally thank D.~Fournier for providing the simulation data used in Fig.~\ref{fig:critical-latitude}, and Z.-C.~Liang and Y.~Mandowara for providing the observational data shown in Fig.~\ref{fig:gizon21_comp}.
Y.B. thanks the organizers of the SDO 2025 Science Workshop (Boulder, 17–21 February 2025) for the invitation to present a plenary talk that motivated this review article. We also benefited from stimulating scientific discussions at the annual WHOLESUN workshops and at the  International Workshop on Low-Frequency Oscillations of the Sun and Stars (Berlin, 15–19 December 2025).  
\end{acks}
 
\begin{authorcontribution}
Y.B developed the structure of the review, prepared all the figures, and wrote the initial draft of the manuscript. Both authors contributed to revising and finalizing the manuscript.
\end{authorcontribution}

\begin{fundinginformation}
Financial support from the ERC Synergy Grant WHOLESUN (grant agreement No. 810218) to L.G. is gratefully acknowledged.
\end{fundinginformation}

\begin{dataavailability}
The data underlying Figs.~\ref{fig:uniformrotation}, \ref{fig:diffrot}, and \ref{fig:gizon21_comp} were newly generated using the code and setups described in the text. These data are available from the corresponding author upon reasonable request.
\end{dataavailability}



\begin{ethics}
\begin{conflict}
The authors declare that they have no conflicts of interest.
\end{conflict}
\end{ethics}

\bibliographystyle{spr-mp-sola} 
\bibliography{ref} 

\end{document}